\documentclass[twocolumn,showpacs,preprintnumbers,amsmath,amssymb]{revtex4}

\usepackage{epsfig}
\usepackage{dcolumn}
\usepackage{bm}

\begin{document}

\title
 {Shell-model description of N$\simeq$Z  $1f_{7/2}$ nuclei}
\author
 {F.~Brandolini and 
  C.A.~Ur\footnote[1]{On leave from NIPNE, Bucharest, Romania}}
\affiliation
 {Dipartimento di Fisica dell'Universit\`a and INFN Sezione di Padova,
  I--35131 Padova, Italy}

\date{\today}

\begin{abstract}

The available  experimental spectroscopic data for nuclei in the middle and in
the second half of the  1$f_{7/2}$ shell are well reproduced by shell model
calculations. For natural parity states of several odd--$A$ nuclei a comparison of shell model calculations in the full $pf$ configuration space 
 with the Nilsson diagram and particle--rotor
predictions shows that prolate strong coupling applies at low excitation energy,
revealing multi--quasiparticle rotational bands and, in some cases, bandcrossings.
Rotational alignment effects are 
observed only in nuclei at the beginning of the shell. 
Moreover, ground state bands experience a change from collective to 
non--collective regime, approaching the termination in the 1$f_{7/2}^n$ space.
 Similar features are observed in the even--even $N$=$Z$ nuclei $^{48}$Cr and $^{52}$Fe
and $N$=$Z$+2 nuclei $^{46}$Ti and $^{50}$Cr. In the $N$=$Z$ nuclei evidence of
the vibrational $\gamma$--band is found. 
 A  review of non natural parity structures is furthermore presented.

\end{abstract}


\maketitle

\section{Introduction}

During the last decade an extensive theoretical work has been made to improve 
the quality of shell model (SM) description of the 1$f_{7/2}$ nuclei, with
particular care to understand the origin of their rotational
collectivity~\cite{Caur1}. The first nuclei studied in detail were
$^{48}$Cr~\cite{Caur2} and $^{50}$Cr~\cite{Mart1}, followed by several odd--$A$ nuclei. One mentions in particular  the 
mirror pair $^{49}$Cr--$^{49}$Mn, whose interpretation was mediated by the predictions of the particle  rotor model (PRM)~\cite{Mart2}.
 
These theoretical advances  were accompanied by a parallel experimental work at the National Laboratories of Legnaro (LNL), that exploited the
advantages of the large $\gamma$--detector array GASP. The level schemes of
$^{48}$Cr and $^{50}$Cr have been extended~\cite{Lenz1,Lenz2} and the
understanding of their collective properties was greatly increased with the 
help of lifetime measurements~\cite{Bra48CR50}.
The research was thereafter extended to several nuclei in the middle of 
the 1$f_{7/2}$, where the attention was focused on the yrast sequence of levels up to the smooth band termination in the 1$f_{7/2}^n$ and 1$d_{3/2}^{-1}\otimes1f_{7/2}^{n+1}$
configuration spaces, for natural parity and unnatural parity, respectively. 
These structures were efficiently populated in heavy--ion induced fusion reactions.
 SM calculations in the full {\it pf} configuration space~\cite{Caur1} reproduce very well the excitation energies of the observed natural parity levels and the $B(E2)$ and $B(M1)$ rates. The double shell closure at $^{56}$Ni is strong 
enough to produce clear smooth terminations in the 1$f_{7/2}$ shell, but weak enough to allow for a large mixing with the other orbitals of the {\it pf} 
configuration space, giving rise to  collective effects which could hardly be 
imagined to occur in this nuclear region.

In general, low-lying levels can be classified in the framework of the prolate Nilsson diagram.
The crossing of the ground state (gs) band with a sideband was first observed 
in $^{50}$Cr. The sideband was interpreted as a 4--qp $K^\pi$=10$^+$ band, 
originating from the simultaneous  excitation of a proton from the [321]3/2$^-$
orbital to the [312]5/2$^-$ one and a neutron from the [312]5/2$^-$ orbital to
the [303]7/2$^-$ one~\cite{BraSev}.
In a recent paper experimental evidence of the two  2--qp bands
with $K^\pi$=4$^+$ for protons and $K^\pi$=6$^+$ for neutrons was  
found~\cite{Bra50CR}. It was also shown   that  the observed mirror energy
differences~\cite{Lenz3}, mainly due to Coulomb energy differences
(CED), are consistent with deformation alignment (strong coupling).

 SM~\cite{Caur2}, cranked Hartree--Fock--Bogoliubov
(CHFB)~\cite{Caur2,Tanaka} and cranked Nilsson--Strutinsky (CNS)~\cite{Juo2}
calculations do not confirm the origin of the backbending at $I^\pi=12^+$ 
 in $^{48}$Cr as due to the bandcrossing inferred by projected shell model (PSM) calculations~\cite{Ha}.
It has been rather related to the smooth termination in a  $v$=4 seniority
subspace~\cite{BraSev}, as suggested by the calculations of Ref.~\cite{Juo1}. 
 
In $^{49}$Cr, the backbending of the $K^\pi$=5/2$^-$ gs band at 19/2$^-$ was
similarly interpreted as a smooth termination in a $v$=3 
subspace~\cite{BraSev}.
The level at 3528 keV was identified as the head of a 3--qp $K^\pi$=13/2$^-$
band, which is described as due to the excitation of a proton from the
[321]3/2$^-$ to the [312]5/2$^-$ orbital followed by the coupling to the 
maximum $K$--value of all unpaired nucleons~\cite{Bra49CR,Bra47V49CR}. Recently, 
two more members of this band were observed, as well as states belonging to 
the 1-qp bands based on the Nilsson orbitals [321]1/2$^-$, [303]7/2$^-$ and [321]3/2$^-$~\cite{BraUNP}.
Very interesting features were observed also for the positive parity levels.
Above the 1-qp $K^\pi$=3/2$^+$ band built on 
the [202]3/2$^+$ orbital, a 3--qp $K^\pi$=13/2$^+$ band becomes yrast and acts to trap the decay flux towards positive parity levels of the $K^\pi$=3/2$^+$ band.
The 3--qp band is described as due to the excitation of a proton from the
[202]3/2$^+$ orbital to the empty [312]5/2$^-$ one, followed by the coupling to the
maximum $K$ of the three unpaired nucleons~\cite{Bra49CR} or, in a different 
representation, as a  $\pi d_{3/2}^{-1}\otimes ^{50}$Mn(I$=$5,T$=$0)
configuration. 

   

In this work we shall mainly discuss natural parity bands in several $N$$\simeq$Z 
odd--$A$ and even-even nuclei pointing out further  evidences of 
smooth terminations, multi-qp bands, bandcrossings and non--collective regime, 
 not considered in detail in  recent reviews \cite{Caurlast,KB3G} and complementing  the pioneering work of Ref.~\cite{Mart2}.
For this purpose the calculated static electromagnetic (em) moments provide 
selective probes of the underlying structure of the nucleus.
 It will be assumed that these observables are fully reliable, unless
specifically mentioned. This is based on the good agreement achieved for level
schemes and  $B(E2)$ and $B(M1)$ values in 
 most discussed nuclei and in particular in $^{49}$Cr,
$^{48}$Cr and $^{50}$Cr~\cite{Bra47V49CR,Bra48CR50,BraSev,Bra50CR}, which are discussed here. 

 For reasons of clarity, the comparison will be restricted to levels up to the termination in the 1$f_{7/2}^n$ configuration space even if good agreement was found also above 
it~\cite{KB3G}. For the same reason odd-odd nuclei will not be  discussed in spite of excellent agreement achieved \cite{Bra46V,Bra48V}.

\section{About shell model calculations}
 
 SM calculations will be compared with experimental level schemes taken from the last
review in Nuclear Data Sheets (NDS) and  most recent references. 
 
Calculations for the natural parity levels have been made in the full {\it pf}
shell using the effective interactions KB3~\cite{Caur1}, KB3G~\cite{KB3G} and
FPD6~\cite{FPD6}. The KB3G interaction is just a slight modification of KB3 to
better reproduce the nuclear properties  approaching $^{56}$Ni. Generally, their
results are rather similar but in  particular cases some gives better
predictions.
For instance, in proximity of $^{48}$Cr the KB3/KB3G interaction gives  excellent
predictions while  FPD6  is about ten percent too strong. 
Recently, the GXPF1 interaction has been developed for nuclei around and 
above $^{56}$Ni~\cite{Ots,Hon}. This interaction is rather equivalent to  KB3G 
 for the nuclei considered here but the energies are somewhat worse 
reproduced. This is why the KB3G interaction was adopted in our calculations. 
The effective interaction is made of a monopole and a multipole part. The 
latter one, which contains most of the structural information, is dominated by 
the pairing and the quadrupole terms. Calculations  considering 
only these two multipole contributions are still made~\cite{Hase}.
 Present calculations were made assuming isospin conservation, i.e. neglecting the effect of Coulomb interaction.  Bare nucleon $g$--factor
values and effective charges (1.5 for protons and 0.5 for neutrons) were 
adopted.
Standard PC computers with Pentium 4 CPU's and 2 GB RAM were used.     The SM code ANTOINE was used, in its WEB distributed version~\cite{ANT1,ANT2}.

Some remarks on the consequences of  configuration space truncation are
worth to be mentioned here. It was observed that most of the rotational
collectivity is already reproduced in a $1f_{7/2}2p_{3/2}$ subspace, where a
``quasi SU(3)'' scheme is valid. This means that the quadrupole deformation originates in a rather similar way as in the SU(3) scheme in the $sd$ shell~\cite{QSU3}. From an
empirical point of view, the main consequences of using the reduced space are
that deformation is somewhat reduced and the 
binding energies of the lower levels are smaller so that their spacing is not 
fully rotational. 
     When truncation is necessary, a limited number of nucleons $s$ is allowed to 
be moved from the $1f_{7/2}2p_{3/2}$ subspace, to the $1f_{5/2}$ and $2p_{1/2}$ 
orbitals. Space truncation was necessary for $A\ge$42.
This truncation differs somewhat from the one commonly adopted, which counts 
all the particles outside the $f_{7/2}$ orbital.
     


\section{Comparison of SM and rotor estimates}

The SM results for em moments will be compared with the predictions of the axial rotor~\cite{BM}, where the configuration is  described by the
Nilsson diagram. The intrinsic electric quadrupole moment, $Q_\circ$, is 
derived from the calculated spectroscopic quadrupole moment, according to the 
formula:

\begin{equation}
 Q_s=Q_\circ\frac{3K^2-I(I+1)}{(I+1)(2I+3)}
 \label{eq01}
\end{equation}

The intrinsic quadrupole moment can also be derived from the $B(E2)$ values using
the formula:
 
\begin{equation}
 B(E2)=\frac{5}{16\pi}Q_t^2<I_iK20|I_fK>^2
 \label{eq02}
\end{equation}
 
\noindent where it is denoted as $Q_t$. 

In the following we will assume $Q_t= Q_\circ$. Its relation with the deformation 
parameter $\beta$ is given by the equation~\cite{Lo}:

\begin{equation}
 Q_\circ=1.09 Z A^{2/3}\beta(1+0.36\beta)\, {\rm fm}^2
 \label{eq03}
\end{equation}
 
For further use, we define the parameter $\beta^*= \beta(1+0.36\beta)$. The reason for this is that in literature some authors adopt the coefficient 0.16 for $\beta$ inside the brackets.

Concerning the magnetic properties of odd-$A$ nuclei, the $g$--factor values are expressed as:

\begin{equation}
 g=g_R + (g_K-g_R) \frac{K^2}{I(I+1)}
 \label{eq04}
\end{equation}
\noindent and the M1 reduced probabilities are:
 
\begin{eqnarray}
 B(M1)= \frac{3}{4 \pi} <I_iK10|I_fK>^2 (g_K-g_R)^2 K^2 \mu_N^2\\
 \label{eq05}
 (for \, K \ne \, 1/2) \nonumber
\end{eqnarray}

In the case of eq.~(5) a decoupling term must be added  when $K=$1/2. 
These formulas have to be considered as qualitative since, even in the extreme 
hypothesis of no residual interaction, $K$-values are mixed by the 
Coriolis force. This is accounted by  
PRM~\cite{Rag}, but it will be shown that, generally, the mixing caused by 
the Coriolis force  is not large in the second half of 
the 1$f_{7/2}$ shell, so that they provide an adequate reference. It is 
necessary to use PRM in the first half of the shell, where the 
[330]1/2$^-$ Nilsson orbital is important and thus the Coriolis force gives rise to a 
partial Coriolis decoupling (CD). 

Eq.~(\ref{eq01}) and (\ref{eq02}) will be applied to even--even nuclei as well.
The magnetic properties of the $N$=$Z$ nuclei $^{48}$Cr and $^{52}$Fe will not be
considered here because they are only slightly sensitive to the nuclear structure: in
self--conjugated nuclei the $g$--factor values approach closely the isoscalar 
value $\displaystyle\frac{g_p+g_n}{2}$, where the nucleon $g$--factors are the 
Schmidt values in the 1$f_{7/2}$ shell, while $B(M1)$ values are very small. On 
the other hand, the magnetic properties of $N$=$Z$+2 nuclei have not a simple 
interpretation.

As previously commented, $K$ cannot be considered a good quantum number. In fact  mixings of the order of ten percent are rather often estimated, usually larger in the full {\it pf} configuration space  than in the restricted $f_{7/2}p_{3/2}$ one. This  makes the 
observation of intraband transitions  difficult, because they are unfavoured 
by the low transition energy and it may perturb the values of em moments but, hopefully, without obscuring the basic structural effects.
 The mixing increases with  spin due to both  the decrease of the deformation and  the higher level density.
 
 The  1-qp sidebands will be not discussed in
detail, since experimental data are often not precise. The  Nilsson
orbital assignments in the displayed level schemes are in general tentative.

\subsection{$^{53}$Fe and $^{43}$Sc}

\begin{figure}[h]
 \epsfig{file=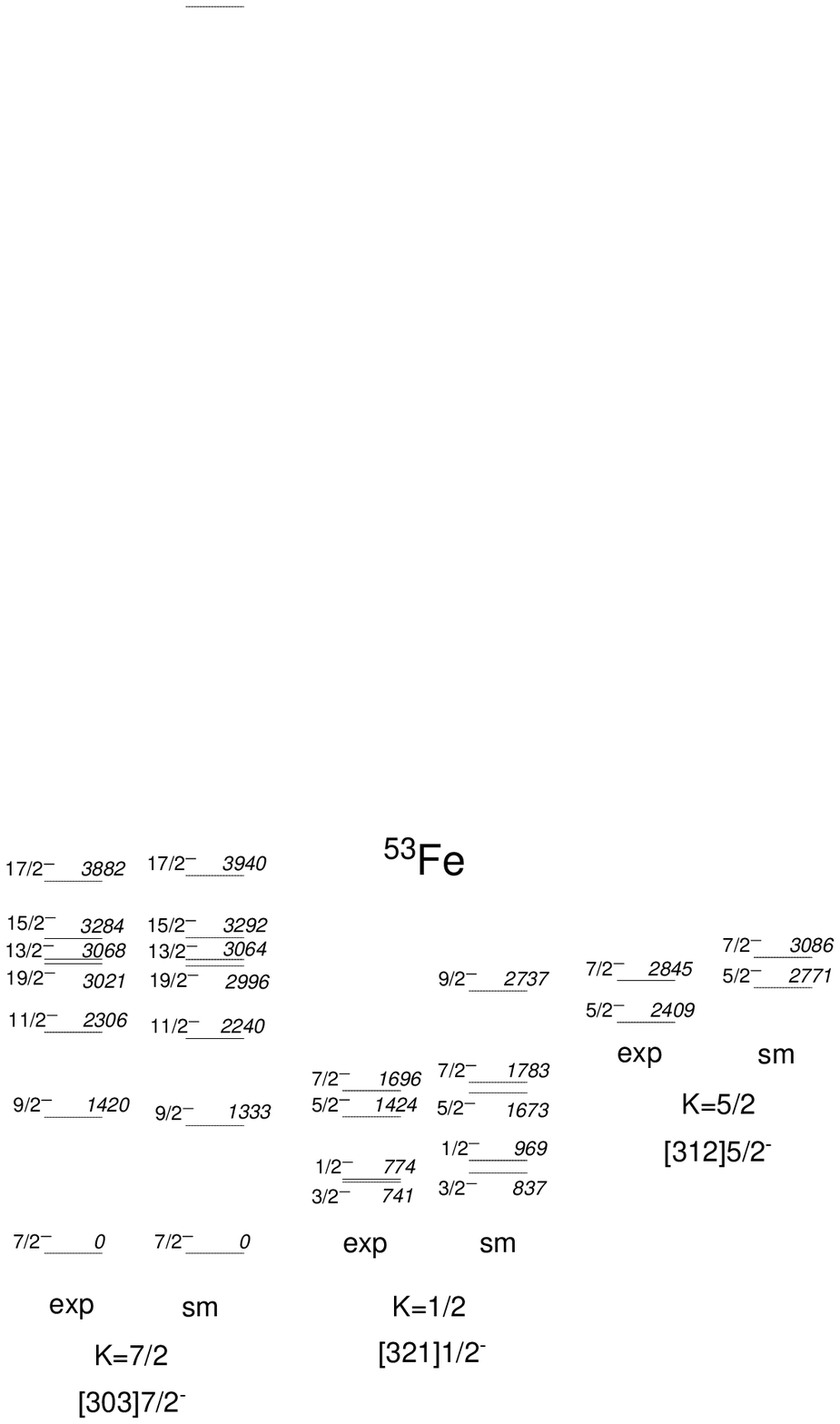,height=8cm ,width=8.5cm,angle=0}
 \protect\caption{Comparison of experimental negative parity levels in $^{53}$Fe with 
                  SM predictions ($s=$3). }
 \label{fig01}
 \epsfig{file=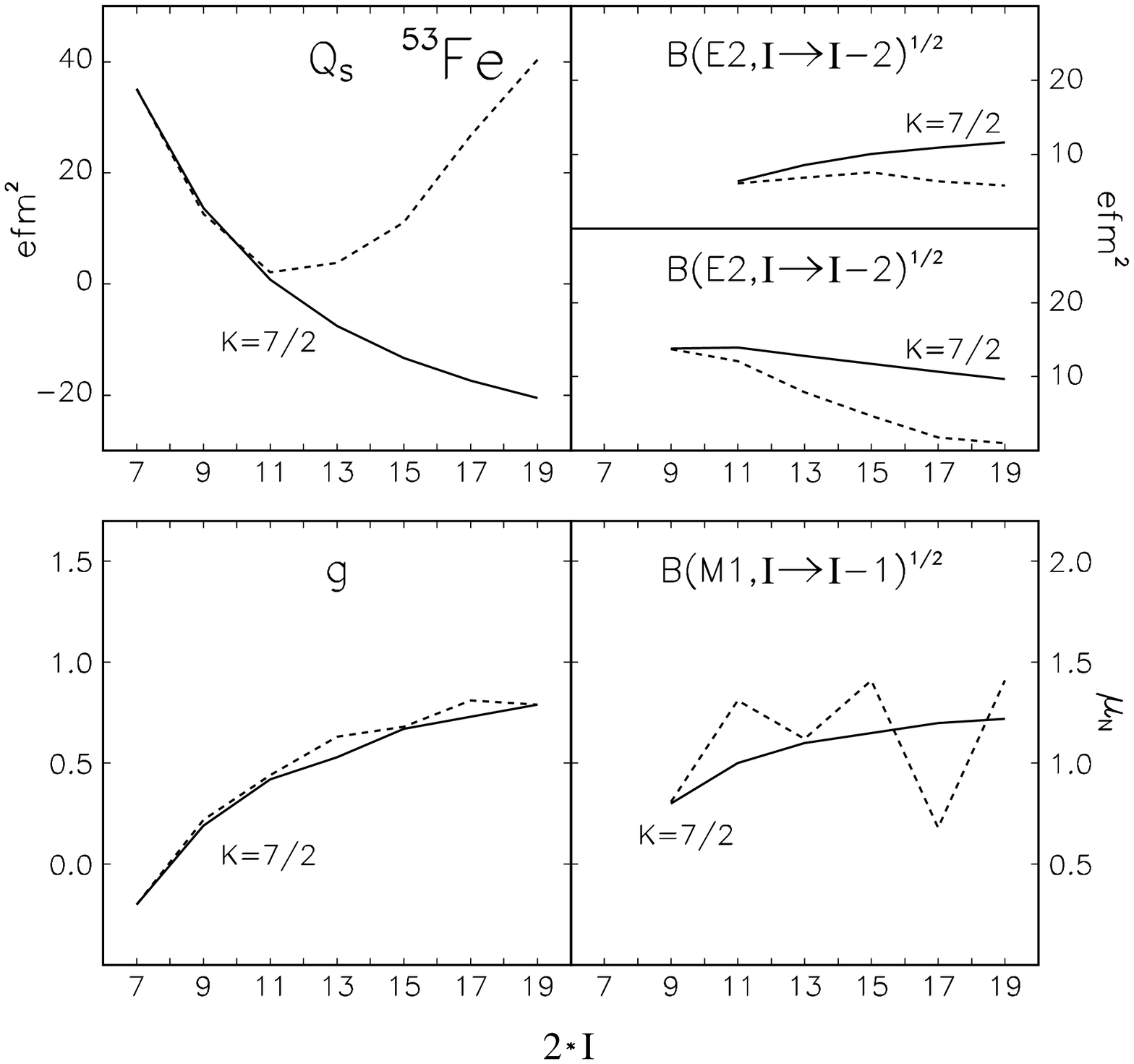,width=8.cm,angle=0}
 \protect\caption{Calculated em moments in the gs band of $^{53}$Fe.
                  Rotor ($\beta^*$=0.19): solid lines, SM: dashed lines.  }
 \label{fig02}
\end{figure}

For our scope, it is important to examine in detail the gs band of the
nucleus $^{53}$Fe, whose terminating state ($I^\pi=19/2^-$, $E_x$=3049 keV, $\tau=$3.6 min), is dominated by a 
$\pi f_{7/2}^{-2}(I=6) \otimes \nu f_{7/2}^{-1}$ configuration and it is  an
yrast trap, as it lies four hundred keV below the yrast 15/2$^-$ state. Its
experimental level scheme is compared with the SM one in Fig.~\ref{fig01},
where a $s=$3 truncation was adopted. Experimental levels were taken from NDS and a
recent paper~\cite{Will}. Two 1-qp sidebands are tentatively assigned. The largest contribution to the wavefunction of the yrast 1/2$^-$  and 3/2$^-$ levels
comes from the excitation of one neutron to the $2p_{3/2}$ orbital, as expected in the case of the [321]1/2$^-$ orbital.
  

$^{53}$Fe is a very convenient testing bench to investigate how rotational
collectivity builds up. In the second half of the 1$f_{7/2}$ shell there is no
disturbance from 2- and  4-hole configurations which are effective at the beginning
of the shell~\cite{Ger} and moreover the $2p_{3/2}$ orbital is very active in
generating deformation. $^{53}$Fe is predicted to be rather deformed in low 
lying states. Experimentally, this is shown by the observed fast E2 (31 Wu) 
transition from 5/2$^-$ to 1/2$^-$ which belongs to the K$=$1/2 sideband.

\begin{table*}
 \caption{\label{tab01}$^{43}$Sc.}
 \begin{ruledtabular}
 \begin{tabular}{c|c|c|c|c|c|c|c|c|c} 
 I & $E_x$ & $E_{sm}$& $Q_s$ & $B(E2,\Delta$I=1) & $B(E2,\Delta$I=2)& $E_{prm}$ & $Q_s$ & $B(E2,\Delta$I=1) & $B(E2,\Delta$I=2)  \\
 &keV &keV &efm$^2$ &e$^2$fm$^4$ &e$^2$fm$^4$    & keV  &efm$^2$ &e$^2$fm$^4$ &e$^2$fm$^4$ \\	  
 \hline
 ~$7/2^-$ &    0       &  0  &-18 &  -  &  -   &   0      &   -8 & -   &  -   \\
 ~$9/2^-$ & 1883 &1999 & -11 & 14 &    -   & 1905 & -27 &14 &  -   \\
 $11/2^-$ & 1830 &1816 & -17 & 16 & 26& 1570 & -13 &  8 & 27 \\
 $13/2^-$ & 3958\footnotemark&3445 & -10 & 5 & 13 & 5364 & -9 & 8 & 32 \\
 \end{tabular}
 \end{ruledtabular}
 \footnotetext{Ref.~\cite{Esp}}
\end{table*}
In Fig.~\ref{fig02} all calculated em observables in the $^{53}$Fe gs band are
compared with the predictions for a prolate rotor.
Differently from Ref.~\cite{Mart2}, $Q_s$ rather than $Q_\circ$ is plotted since it will be shown that eq.~(\ref{eq01}) cannot be applied 
in general. Furthermore, the square root of the $B(E2)$ values are reported 
because they are approximately proportional to the deformation as the  $Q_s$ values 
do. The  $Q_s$ value of the gs state is predicted to be large and positive, as
expected  for a prolate $K$=7/2 band with $\beta^*\simeq$ 0.19.
Calculated $Q_s$ and $B(E2)$ values agree with rotor predictions only at low spins 
 up to 11/2, while  the typical  $I(I+1$) spectrum has not room to get 
evident. What occurs above 11/2 will be discussed later.
In the 1$f_{7/2}^n$ space cross--conjugate nuclei should have identical level 
scheme, while, experimentally, large difference are observed, pointing to
different configuration mixing and collective properties. The predicted level
scheme for the $\pi^{-2}\nu^{-1}$ $^{53}$Fe is in fact substantially different
from that of its cross--conjugate $\pi^1\nu^2$ $^{43}$Sc. In both cases the gs
level is 7/2$^-$, but in $^{43}$Sc a large signature splitting (SS) is observed.
Another striking difference is that the $B(E2)$ values of the transition from 
the yrast level 9/2$^-$ to the 7/2$^-$ in $^{43}$Sc and $^{53}$Fe are 13.6 and 217.2 efm$^2$, respectively.
The latter is a fingerprint of the  prolate collectivity in $^{53}$Fe, since in
a 1$f_{7/2}^3$ space only one third of this value is predicted. 
The properties of $^{43}$Sc, on the other side, can be roughly described by PRM
predictions, assuming a slightly prolate shape  with $\beta^*$=0.10, which predict
rotational alignment in the gs band built on the [330]1/2$^-$ Nilsson  orbital. 
SM predictions are compared with PRM predictions in Table~\ref{tab01}. There is 
some correspondence, in spite of the naive description. It has to be remarked,
however that  SM calculations are not good in $^{43}$Sc, because  2h-- and 
4h--configurations strongly mix at low spin in nuclei close to 
$^{40}$Ca~\cite{Ger}. This does not occur in the isotope $^{47}$Sc with
$\pi^1\nu^{-2}$, where good agreement is achieved by SM~\cite{Mart2}. Both $^{43}$Sc and $^{53}$Fe, as
in general 1$f_{7/2}$ nuclei,  are described with  a prolate shape.

The different structure of the cross--conjugate nuclei is confirmed by the
contribution of the 1$f_{7/2}^n$ configuration (expressed in percentage) to the  states along the gs band.
For the levels with spin  7/2, 9/2, 11/2, 13/2, 15/2, 17/2 and 19/2 the contributions are 79, 96, 72, 95, 78, 88  and 96 in $^{43}$Sc and 56, 54, 58, 60, 63, 60 and 64 in $^{53}$Fe.
 The staggering of the values in $^{43}$Sc is related to the observed SS, where the 
7/2$^-$ ground state has the favoured signature.
These numbers provide, moreover, a qualitative explanation for the inversion 
between the 15/2$^-$ and 19/2$^-$ observed in $^{53}$Fe.
The level scheme calculated in the 1$f_{7/2}^{\pm 3}$ space predicts the inversion, 
as a consequence of the very attractive $V_{pn}(I=7,T=0)$ term. The inversion is 
preserved in $^{53}$Fe because the contribution of 1$f_{7/2}^{-3}$ is 
similar in the two states, while in $^{43}$Sc the 1$f_{7/2}^{3}$ contribution is about 20 \% larger for the 19/2$^-$, so that the 15/2$^-$ has an additional binding energy, that 
lowers it below the 19/2$^-$  level. 
  More than one third of the upper orbital occupation refers to the $2p_{3/2}$ one, which give rise to large quadrupole terms.
  
CD is expected to be large at low spin if the gs band is based on the [330]1/2 
orbital as in $^{43}$Sc, it is noticeable for the [321]3/2$^-$  orbital,  giving rise to partial rotational alignment (RAL), and 
nearly negligible for the  orbitals [312]5/2$^-$ and [303]7/2$^-$.  It  becomes 
important  in the case of the intruder orbital [440]1/2$^+$, originating from 
the spherical 1$g_{9/2}$ orbital.

As shown in Fig.~\ref{fig02}, the $Q_s$ values of the 
$^{53}$Fe gs band start to increase above $I^\pi=$11/2$^-$,  reaching at the terminating  spin 19/2$^-$ a maximum of  42.9 efm$^2$, calculated with $s$=4. This value may be somewhat larger
in full $pf$ calculations. This is consistent with its hole--like nature, to 
which a prolate non--collective shape pertains, being made of three valence 
nucleon--hole. In a semiclassical description $Q_s=Q_\circ$ and thus 
$\beta^*\simeq 0.11$ is derived from eq.~(\ref{eq03}). 
Since $Q_s=25$ efm$^2$ is predicted in the 1$f_{7/2}^n$ space, one infers that
the terminating state is strongly polarised by the upper orbitals, getting a
strong enhancement of the non--collective prolateness. Its terminating nature is
confirmed by the fact that yrast 21/2$^-$ and 23/2$^-$ levels are predicted more than 4 MeV above it.  
The $Q_s$ value is stable upon large variations of the binding energies of 
the upper orbitals.
It would be of great interest to measure the $Q_s$ value of the 19/2$^-$ terminating state.

In a  $\beta-\gamma$ plane the shape changes from $\gamma\simeq 0$ at the bottom
of the gs band to $\gamma\simeq -120$ at its termination (Lund convention).
While it is commonly accepted that the terminating states are non-rotational, 
it is not trivial to explain how the nuclear shape evolves along the gs band. This problem
has been commonly addressed with the configuration dependent CNS 
approach~\cite{RagNS}, as recently made for $^{50}$Mn, where the low-lying bands
terminate as prolate non--collective \cite{Guo}.
One may question whether a meanfield description as that of CNS can describe 
accurately a system with few valence particles or holes. One should find that 
in few steps the nucleus  changes in sequence to triaxial collective, oblate 
collective and finally prolate non--collective shapes. It seems more
realistic to imagine a rapid change from prolate collective to prolate 
non--collective, because an oblate collective shape is likely inhibited by the
simple shell model structure. On the other hand, it is known since a long time 
that the single particle structure of a prolate nucleus may favour a prolate 
non--collective shape~\cite{Aberg}.
More arguments for a sudden shape change will be presented in the discussion 
of other nuclei.
  
The evolution of $Q_s$ values approaching the terminating spin 19/2$^-$ is 
certainly incompatible with rotational alignment, elsewhere  proposed~\cite{Sh},
because a negative value of $Q_s$ would be expected in that case, as one has to 
put a large value of $I$ and a small one for $K$ in eq.~(\ref{eq01}).

In the past, the non--collective prolate shape at the band termination was alternatively interpreted as 
a high--$K$ collective prolate shape~\cite{Ur}. This misunderstanding originates  from the large overlap of the two wave functions: in fact a $I$=$K$=19/2 state is 
oriented spacially almost in the same manner, where $Q_s$=0.74$Q_\circ$, according 
to eq.~(\ref{eq01}). The substantial physical difference is that the 
non--rotational state is a terminating one, while the hypothetical $K$=19/2 state 
would be a bandhead. The latter description is clearly excluded by the 
observation that  $Q_s$ increases gradually with the level spin so that also previous levels should 
have a dominant high--$K$ character, which is unrealistic.

\begin{table}[b]
 \caption{\label{tab02}$Q_s$ and $g$--factor values of terminating states.}
 \begin{ruledtabular}
 \begin{tabular}{c|c|c|c|c|c|c|c} 
 Nuclide&config.&$I^\pi$&P($f_{7/2}$)&$Q_s$&$Q_s$($f_{7/2}$)&g$_{sm}$&g$_{emp}$\\
 & & & \% &efm$^2$ &efm$^2$ & & \\
 \hline
 $^{55}$Co&$\pi^{-1}        $&$ 7/2^-$&62.6& 24.1  & 17.5& 1.350    &1.38\\
 $^{55}$Ni&$\nu^{-1}        $&$ 7/2^-$&   "& 14.8&  5.8& -0.271    &-0.30\\
 $^{54}$Fe&$\pi^{-2}        $&$ 6^+  $&61.1& 27.2& 19.8& 1.337    &1.38\\
 $^{54}$Ni&$\nu^{-2}        $&$ 6^+  $&   "& 18.9&  6.6& -0.256    &-0.30\\
 $^{53}$Fe&$\pi^{-2}\nu^{-1}$&$19/2^-$&60.7& 42.9& 25.5&0.770&0.80\\
 $^{53}$Co&$\pi^{-1}\nu^{-2}$&$19/2^-$&   "& 43.1& 23.8&0.318&0.28\\
 $^{52}$Fe&$\pi^{-2}\nu^{-2}$&$12^+  $&60.9& 47.1& 26.1& 0.547  &0.54\\
 $^{51}$Mn&$\pi^{-3}\nu^{-2}$&$27/2^-$&68.4& 32.7& 18.6&0.646&0.65\\
 $^{51}$Fe&$\pi^{-2}\nu^{-3}$&$27/2^-$&   "& 39.4& 23.5&0.446&0.43\\
 $^{50}$Cr&$\pi^{-4}\nu^{-2}$&$14^+  $&75.3&  7.1&  6.4&0.668&0.68\\
 $^{50}$Fe&$\pi^{-2}\nu^{-4}$&$14^+  $&   "& 22.3& 19.3&0.421&0.40\\
 $^{49}$Cr&$\pi^{-4}\nu^{-3}$&$31/2^-$&88.9&  1.6&  4.0&0.586&0.57\\
 $^{49}$Mn&$\pi^{-3}\nu^{-4}$&$31/2^-$&   "&  9.8& 12.0&0.503&0.51\\
 $^{48}$Cr&$\pi^{ 4}\nu^{ 4}$&$16^+  $&89.4& -6.7&  0  & 0.553 &0.54\\
 $^{47}$V &$\pi^{ 3}\nu^{ 4}$&$31/2^-$&87.5&-19.2&-12.0&0.534&0.51\\
 $^{47}$Cr&$\pi^{ 4}\nu^{ 3}$&$31/2^-$&   "&-14.4& -4.0&0.549&0.57\\
 $^{46}$Ti&$\pi^{ 2}\nu^{ 4}$&$14^+  $&86.7&-24.2&-19.3&0.449&0.40\\
 $^{46}$Cr&$\pi^{ 4}\nu^{ 2}$&$14^+  $&   "&-18.8& -6.4&0.638&0.68\\
 $^{45}$Ti&$\pi^{ 2}\nu^{ 3}$&$27/2^-$&90.0&-26.8&-23.5&0.471&0.43\\
 $^{45}$V &$\pi^{ 3}\nu^{ 2}$&$27/2^-$&   "&-25.8&-18.6&0.616&0.65\\
 $^{44}$Ti&$\pi^{ 2}\nu^{ 2}$&$12^+  $&94.3&-27.7&-26.1& 0.545 &0.54\\
 $^{43}$Sc&$\pi^{ 1}\nu^{ 2}$&$19/2^-$&96.1&-23.1&-23.8&0.330&0.28\\
 $^{43}$Ti&$\pi^{ 2}\nu^{ 1}$&$19/2^-$&   "&-26.3&-25.5&0.765&0.80\\
 \end{tabular}
 \end{ruledtabular}
\end{table}
 
In Table~\ref{tab02} the predictions of $Q_s$ values for several terminating states in 
$N$$\simeq$$Z$ nuclei are compared with $f_{7/2}^n$ predictions. Calculations were made in the full $pf$ configuration space, except for 52$\le A \le$54 and $A$=55 where  $s$=4 and $s$=3 truncations were assumed, respectively. 
Obviously, in the Table a 4--hole configuration coincides with a 4--particle one.
 These states are generally supposed to have nearly pure $f_{7/2}^n$ configuration,
but this turns out to be not true approaching the end of the shell, where 
configuration mixing becomes important. 
One observes that a negative value of $Q_s$ is calculated for $^{48}$Cr, where 0 is expected in the $1f_{7/2}$ space. Similarly, a negative offset applies to neighbouring nuclei.
 The scale factor of $Q_s$ for the single proton--hole nucleus $^{55}$Co in the
$f_{7/2}^n$ is $(2j-1)/(2j+1) <r^2>$~\cite{BM}.
  It is peculiar that $\nu^{-2}\pi^{-1}$ and $\nu^{-1}\pi^{-2}$ configurations 
have a similar $Q_s$ value.

 The experimental $g$-factor value of the
19/2$^-$ level in $^{43}$Sc is known to be 0.329(1), that of the 6$^+$ level in $^{54}$Fe is  1.37(3) and finally the $g$-factor of the single hole 7/2$^-$ level in $^{55}$Co is 1.378(1). Thus, all  experimental values known with high precision agree with the SM predictions of Table II.

 As previously stated, the interaction KB3G has been adopted in Table II.
Calculations made with the GXPF1 one are nearly equivalent, while the  FPD6 
interaction would predict a mixing  about 30\% higher for the terminating 
states approaching shell closure, accompanied by up to 20 \% larger $B(E2)$ 
rates. It is a critical point the capability of the effective interaction to 
get good results in proximity of the shell closure. The experimental value 
B(E2,$6^+$$\rightarrow$$ 4^+$)=39.7(5) efm$^2$ in $^{54}$Fe provides a test for 
the effective interactions. 
The interactions KB3G, GXPF1 and FPD6 gives B(E2,$6^+$$\rightarrow$$4^+$) values 40.3, 
40.7  and 46.7 efm$^2$, using a $s$=4 truncation. The $1f_{7/2}^n$ prediction is 23.1 efm$^2$, 
independently from the adopted interaction. It turns out  that the FPD6 interaction overestimates the $B(E2)$ value by about 20 \%. 


The internal consistence of SM predictions is confirmed  by the $g$--factor 
predictions for the terminating 19/2$^-$ level. It can be estimated with the 
additivity rule:
  
\begin{equation}
 g= \frac{g_p+g_n}{2} + \frac{g_p-g_n}{2} \frac{I_p(I_p+1)-I_n(I_n+1)}{I(I+1)}
 \label{eq06}
\end{equation}
  
\noindent for a pure $\pi f_{7/2}^{-2}(I=6) \otimes \nu f_{7/2}^{-1}$ 
configuration.
If one assumes $g_n$=--0.30 and $g_p$=1.38 in the 1$f_{7/2}$ orbitals, 
instead of the Schmidt values of $-$0.547 and 1.655, respectively, one gets the mean reproduction of SM values of Table II.
These effective values, which roughly account for  configuration mixing,
will be adopted in  semiclassical considerations. In the case 
of the mirror nucleus $^{53}$Co, the isovector part, i.e. the second addendum in eq.~(\ref{eq06}), changes sign. Looking to Table~\ref{tab02} we see that the 
SM isoscalar part agrees with the empirical value of 0.54, calculated as the average of the values for a mirror pair. 

The slope of the $g$ values at low spin follows rather closely a rotor behaviour,
where the asymptotic value at high spins is not the usual $g_R\simeq$0.5, assumed
for a similar and large number of neutrons and protons, but the curve is  
adjusted to g$=$0.77 at $I=$19/2, according to eq.(\ref{eq06}),  because only two proton--holes and one neutron--hole are active.
 The slope of $B(M1)$ is also consistent with a rotor 
behaviour, apart from a staggering approaching termination, which is characteristic 
of a $1f_{7/2}^{-3}$ description.

In summary, $^{53}$Fe can be considered a paradigm of the fragility of the
collective rotation in 1$f_{7/2}$ nuclei and of the intrinsic 
change of regime along the gs band.
 

 
Also $^{43}$Sc is understood in proximity of the termination. There the 
configurations become particle--like, so that a  non--collective oblate 
configuration is expected, to which also a negative $Q_s$ value pertains.
It must be stressed, therefore, that an ambiguity arises for nuclei at the 
beginning of the shell, since  a negative $Q_s$ value is predicted for them both 
before and after the phase transition.


\subsection{$^{49}$Cr}

\begin{figure}[t]
 \epsfig{file=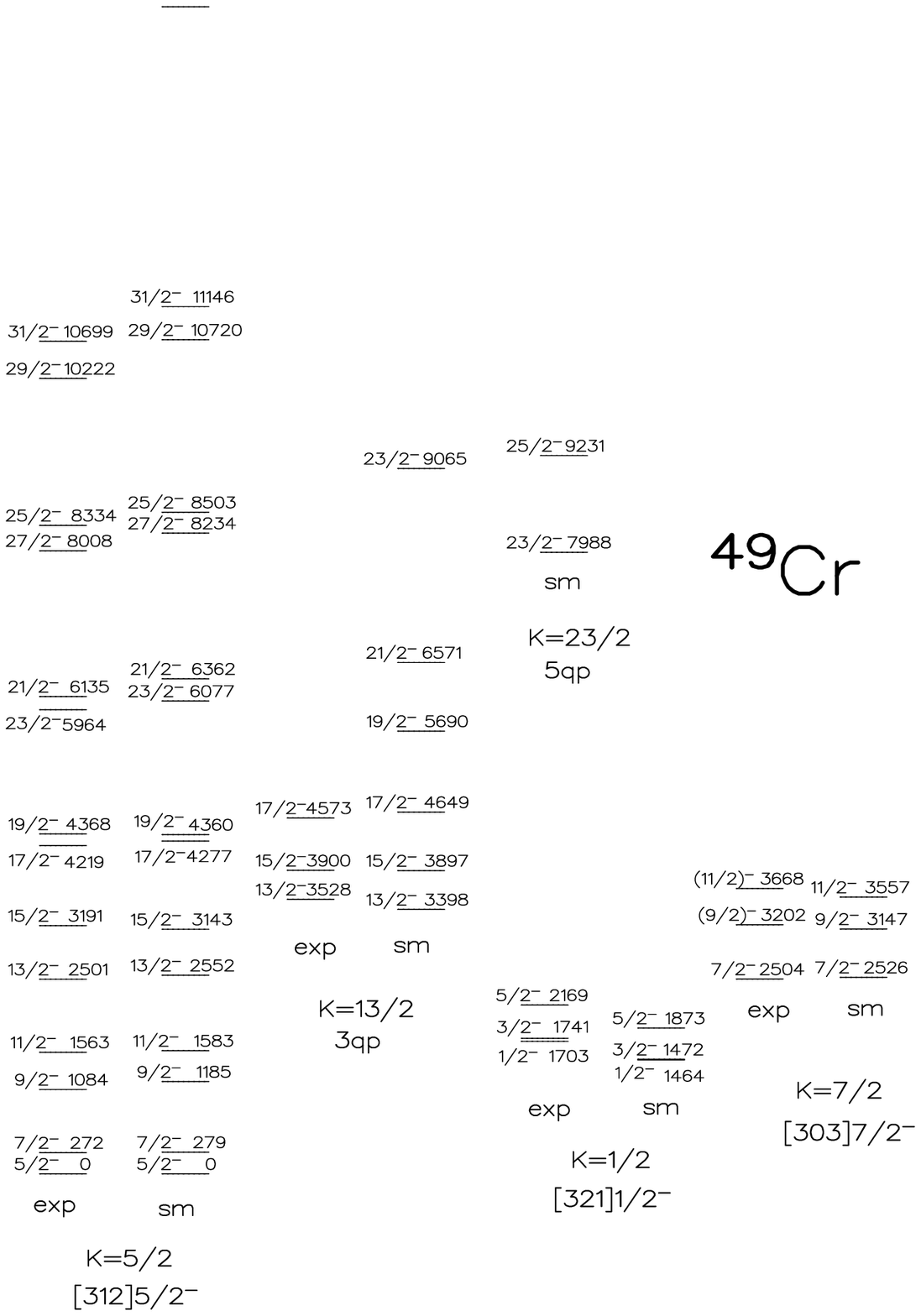,width=8.4cm,angle=0}
 \protect\caption{Comparison of experimental negative parity levels in $^{49}$Cr with SM predictions.}
 \label{fig03}
 \epsfig{file=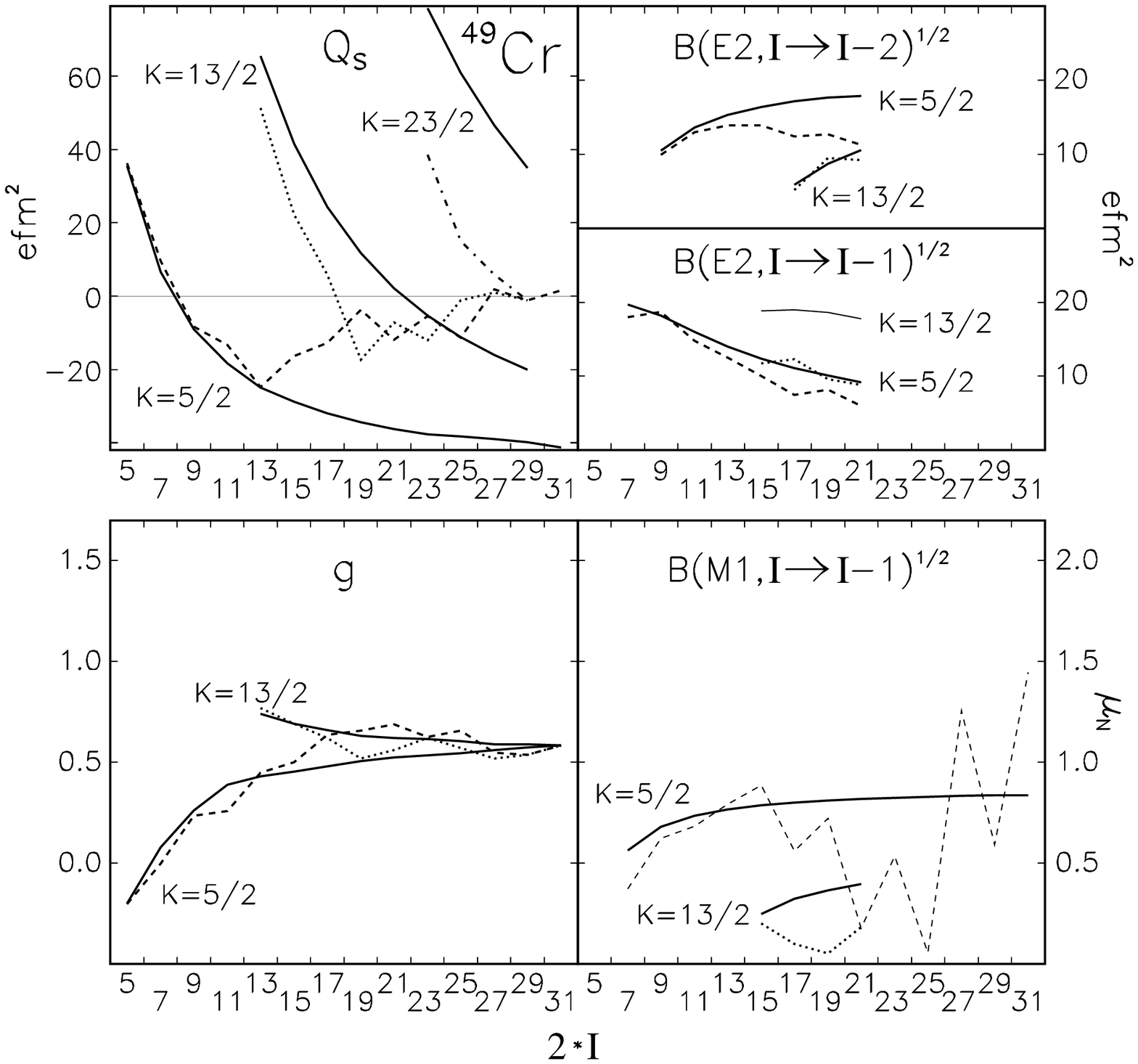,width=8.4cm,angle=0}
 \protect\caption{Calculated em moments in $^{49}$Cr. 
                 Rotor ($\beta^*$=0.28): solid line. SM:  dashed  ($K$=5/2) and  dotted   ($K$=13/2) lines. }
 \label{fig04}
\end{figure}

The next nucleus to be examined is the $N$=$Z$+1 $^{49}$Cr, which has been 
studied in detail in Refs.~\cite{Bra49CR,Bra47V49CR,BraUNP}.
A partial level scheme of $^{49}$Cr is shown in Fig.~\ref{fig03}, up to the 
termination at 31/2$^-$. Only the lower members of the 1-qp sidebands are reported and, approaching the termination, only states with prevaling $1f_{7/2}^n$ configuration. 
  Few levels were added with respect to Refs.~\cite{Bra49CR,Bra47V49CR}, 
which were observed in a recent experimental work~\cite{BraUNP}, where also
a detailed comparison with PRM was made.
Calculated levels  are also organised in bands.

SM calculations are able to reproduce well the observed levels.
The gs band is described at low spins as a $K^\pi$=5/2$^-$ band based on the  
$\nu$[312]5/2$^-$ Nilsson orbital~\cite{Mart2,Bra47V49CR}.
 The 1-qp sidebands with $K$=1/2, 7/2 and 3/2 are   
described with Nilsson orbitals [321]1/2$^-$, [303]7/2$^-$ and [321]3/2$^-$, respectively. Only the first two are reported in Fig.~\ref{fig03}.
A 3--qp band with $K^\pi$=$13/2^-$ is also observed, which is predicted  by breaking a  [321]3/2 proton pair, lifting one of them into the [312]5/2$^-$ and coupling the three unpaired particles to the maximum $K$ value.

The measured lifetimes lead to an initial $\beta^*$ value of 0.28 for the gs band,
which was adopted in Fig.~\ref{fig04} where the em properties are plotted and compared with the rotational  model. 
The $Q_s$ value of the $K^\pi$=13/2$^-$ bandhead is 51.2 efm$^2$, which corresponds to 
$\beta^*=$0.24. The smaller deformation is reasonable owing to the high spin, which reduces the number of possible interacting nucleons in the $2p_{3/2}$ orbital and thus the collectivity. Moreover some mixing is also present.

In the same figures also a 5--qp band with $K^\pi$=23/2$^-$ is reported which is obtained by breaking also a [321]3/2 neutron pair and lifting one of them to the [303]7/2$^-$ orbital. It  is  merely incipient
since it terminates at the common termination 31/2$^-$  ($Q_s=1.6$ efm$^2$).
 The calculated $Q_s$ moment of the $K^\pi$=23/2$^-$ bandhead is 38.8 efm$^2$, corresponding to a $\beta^*=$0.16 value. The yrare 25/2$^-$ state is  mixed so that it decays to the yrast 23/2$^-$ state with a fast M1.

The  magnetic moment of the $I^\pi$=$K^\pi$=$13/2^-$ state is given 
semiclassically by the sum of the longitudinal component along the total spin. In the $1f_{7/2}^n$ space, taking $\cos\theta=\, K/j$ one gets 
$g=\mu/I  \simeq 2/13 [1.38\cdot(3/2+5/2) -0.3(5/2)]$, so that the $g$--factor of the 
13/2$^-$ level is g$=$0.74, in agreement with the SM value of 0.77. The 
curves of $g$ and $B(M1)$ adopt such empirical values.
 It has to be noted that in Fig.~\ref{fig03} some SS is observed at low spin, whose size is not reproduced by PRM and CSM \cite{Beng}.

Above the backbending at 19/2$^-$ the SS of the gs band becomes so large that the 
ordering of spin values is inverted. This indicates  a sudden change of regime, which cannot be associated to a bandcrossing, neither 
experimentally, nor theoretically~\cite{Bra49CR}.

The low spin part  of Fig.~\ref{fig04} is reproduced without triaxiality. 
From the $Q_s$ values it results that the collective rotation of the $^{49}$Cr gs band 
starts to be severely damaged above $I^\pi=13/2^-$. In fact, at low spin they follow the 
rotor predictions getting rapidly negative, according to 
eq.~(\ref{eq01}), but they start to increase around 13/2$^-$, approaching zero at 
$I^\pi=19/2^-$, presumably due to the influence of the $v$=3 termination.
 

This interpretation is confirmed by the $g$--factor curve, adjusted  at $I$=31/2 to 
the empirical value 0.57, obtained from the additivity rule (the SM value is 0.59).
The SM $g$--factor ranges around 0.65 at spin 19/2$^-$, which is 
larger than the rotor value of 0.53, while it is at the halfway to the value of 
0.79 for the termination of $v$=3 states. 

A similar effect  occurs apparently in  $1g_{9/2}^n$ nuclei too. The gs band 
of  $^{93}$Pd, with a configuration $\pi^{-4}\nu^{-3}$, has a backbending at the 
25/2$^+$~\cite{Rusu}, which can be interpreted as a termination in a $v$=3
subspace. Its gs band appears to be little collective, indicating that 
approaching the $N$=$Z$=50 shell closure the polarising action 
of the 2$d_{5/2}$ orbital on the 1$g_{9/2}$ one is less effective than that of 
the 2$p_{3/2}$ orbital on the 1$f_{7/2}$ one, because of the larger energy gap 
between the orbitals.

It appears rather questionable to speak of 
rotational bands above $I\simeq$15/2 since the bands start to  be strongly mixed 
and single particle features get evident. 
 In fact, the staggering of $B(M1)$ values approaching band termination is  a  $1f_{7/2}^n$ phenomenon~\cite{Cam}, which 
 cannot be  reproduced by a deformed meanfield model. 

\subsection{$^{51}$Mn}

\begin{figure}[t]
 \epsfig{file=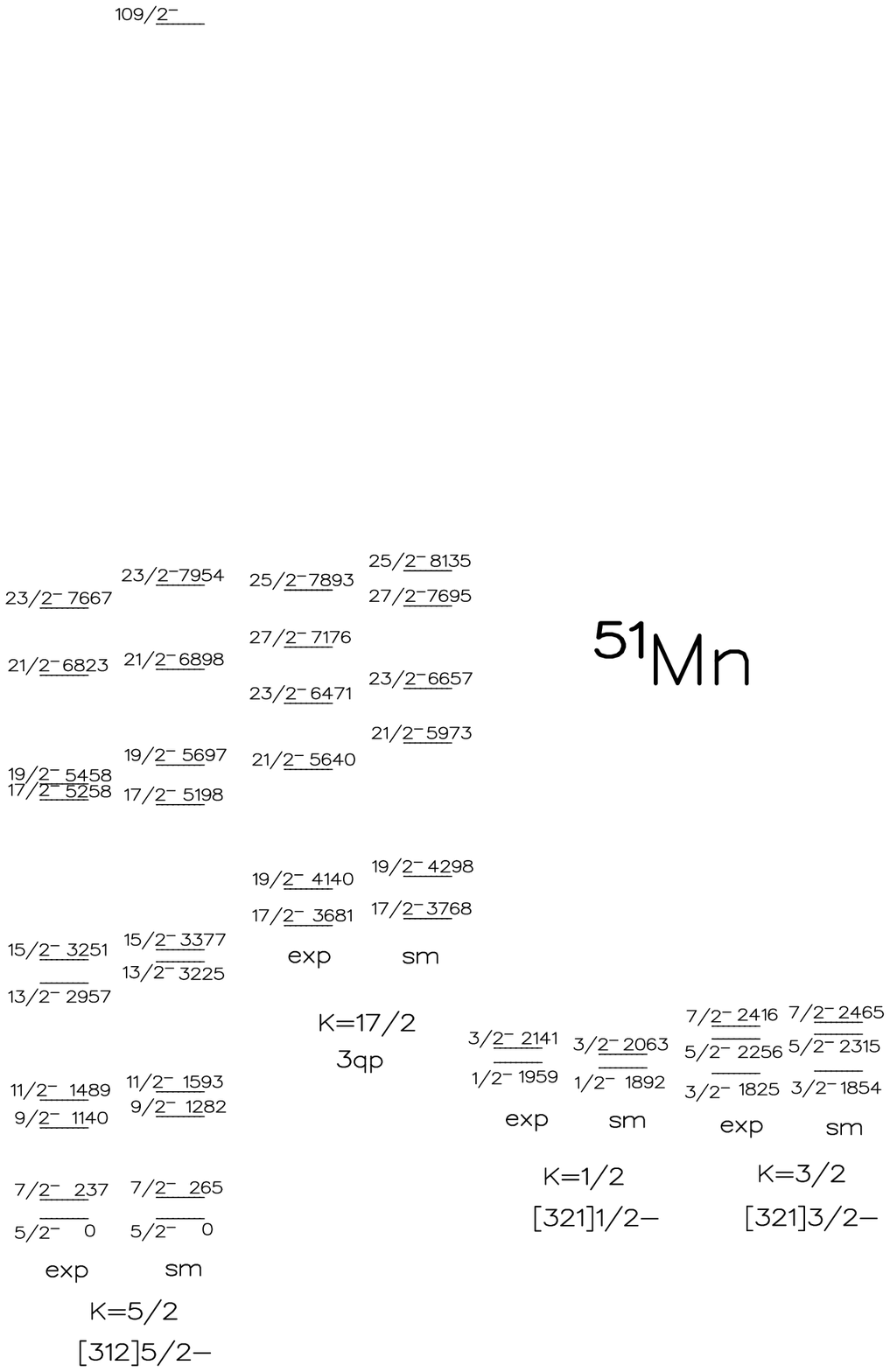,width=8.4cm}
 \protect\caption{Comparison of experimental negative parity levels in $^{51}$Mn with SM predictions.}
 \label{fig05}
 \epsfig{file=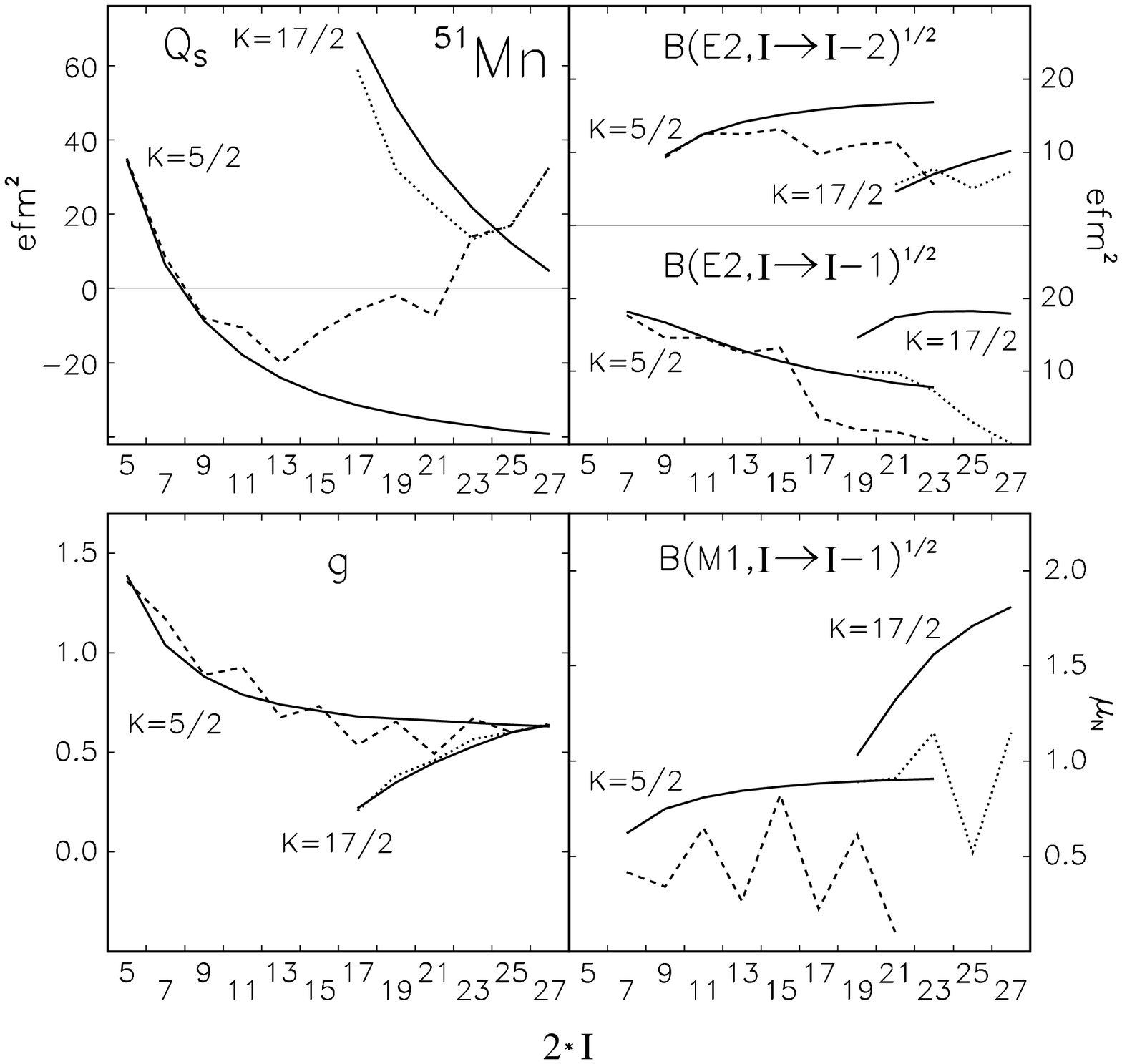,width=8.4cm,angle=0}
 \protect\caption{Calculated em moments in $^{51}$Mn. Rotor ($\beta^*$=0.26): solid lines,  SM: dashed ($K$=5/2) and dotted  ($K$=17/2) lines. }
 \label{fig06}
\end{figure}

Experimental levels, $B(E2)$ and $B(M1)$ values of the $N$=$Z$+1 nucleus $^{51}$Mn have been recently   found to agree with the SM calculations  \cite{KB3G}.
The gs band  is interpreted with the same 
Nilsson configuration [312]5/2$^-$ as $^{49}$Cr, while the termination now 
occurs at 27/2$^-$.
 The experimental level scheme of $^{51}$Mn shown in Fig.~\ref{fig05} includes data from NDS and the very recent Ref.~\cite{Ek2}. 
The $K^\pi$=5/2$^-$ gs band resembles that of $^{49}$Cr, apart from a larger SS,
up to the backbending at 17/2$^-$, pointing to a similar deformation. 
Calculated em observables are plotted in Fig.~\ref{fig06}. 
The larger SS is correlated with the staggering of $B(M1)$ and $g$--factor values. 
CD could produce staggering only at rather high spin, while the observed 
one starts early. 
In  CSM such early staggering  is considered an indication of 
triaxiality, associated with a $\gamma < 0$. The larger SS in $^{51}$Mn with respect to $^{49}$Cr can be interpreted as a sign of substantial triaxiality. 

Staggering was observed also in the CED values~\cite{Bent,Ek1}, which thus may be related with triaxiality.
In spite of that the axial rotor formulas with  $\beta^*$= 0.27  reproduce 
qualitatively  the SM values for $Q_s$ and $B(E2)$ at low spins. 
It has to be noted that above 15/2$^-$ the $B(E2)$ values for $\Delta I=1$ transitions become very small.
 The calculated gs $g$--factor is 1.359 in
agreement with the experimental 1.427 and not far from the effective value for a 
1$f_{7/2}$ proton. 

The 3--qp band is produced by breaking a [312]5/2$^-$ neutron pair, lifting one of them into the [303]7/2$^-$ and coupling the three unpaired particles to the maximum K value.
The interpretation of the yrast 17/2$^-$ level as  a $K^\pi$=17/2$^-$ 
bandhead was already proposed \cite{Bra50CR,BraGASP} to explain the observed anomaly of CED values starting at $I$=17/2. It is confirmed by the predicted large $Q_s$ value of 56.3 
efm$^2$, which corresponds to $\beta^*=0.22$, according to eq.~(\ref{eq01}). 
This estimate may be lowered by  some configuration mixing.
Moreover, the low SM value of its $g$--factor (0.206) is related to the fact that 
now the 3--qp state is formed by  one K$=$5/2 proton and two neutrons with 
K$=$5/2 and 7/2, respectively. Semiclassically, as previously made in 
$^{49}$Cr, $g=\mu/I \simeq 2/17[1.38\cdot 5/2 -0.30(3/2+5/2)]$, resulting in a $g$--factor of 0.26 
for the 17/2$^-$ level.

The calculated yrast level 19/2$^-$ is assigned to the K$=$17/2 band in virtue of its 
em properties.
The $B(E2)$ value of the transition to yrast 
17/2$^-$ level is large, as expected for an intraband transition, while the one
 to the yrast 15/2$^-$ is small.
 Its $g$--factor is 0.39, much smaller than the rotor 
value of 0.60, obtained adjusting the rotor curve to the empirical value of 0.63  at 27/2$^-$ (the SM value is 0.64). 

The yrare 17/2$^-$ level belongs to the gs band since its E2 decay to the yrast 
13/2$^-$ is favoured.
The yrare 19/2$^-$ state was not observed. It is predicted to have g$=$0.67, 
higher than the rotor value of 0.58 and at the halfway to the $v$=3 value 
of 0.79. Its $Q_s$ value is slightly positive.
 
 
The gs band level spacing and $Q_s$ values change rapidly above the 13/2$^-$ level, 
pointing to a change of regime as in $^{49}$Cr, but, while in the case of $^{51}$Mn the SS 
decreases in $^{49}$Cr it increases. The reason of the different behaviour is 
not understood. The change of regime is signalled also by the sudden change of $Q_s$ and $B(E2,I \to I-1)$ values above 17/2$^-$.
       
The positive value of $Q_s$ at the terminating level 27/2$^-$ (32.7 efm$^2$) 
is due to the non--collective prolate shape of the $\pi^{-3}\nu^{-2}$ 
configuration.  The corresponding $\beta^*$ is 0.09.
 The two bands with K$=$5/2 and 17/2 loose  
collectivity with increasing spin and the $Q_s$ value becomes for both positive.

As for $^{53}$Fe, the observed positive $Q_s$ value at 23/2$^-$ and above, it is 
incompatible with rotational alignment, since the $Q_s$ value would be expected to 
be negative in that case.
       
 
Yrast high--$K$ bands of natural parity can occur only in the second half of 
the 1$f_{7/2}$ shell because one needs to deal with high--$K$ Nilsson orbital.
In fact, the reason why the $K^\pi$=17/2$^-$ crosses the gs band, while the $K^\pi$=13/2$^-$ 
band in $^{49}$Cr does not, is that their excitation energy is similar, but the 
spin values are higher in the former case.

Applying the Nilsson diagram to look for a possible 5--qp band as  in 
$^{49}$Cr, one gets $K^\pi$=27/2$^-$, but $I^\pi=27/2^-$ is the terminating state, so 
that a 5--qp band does not exist in $^{51}$Mn.

\begin{figure}[t]
 \epsfig{file=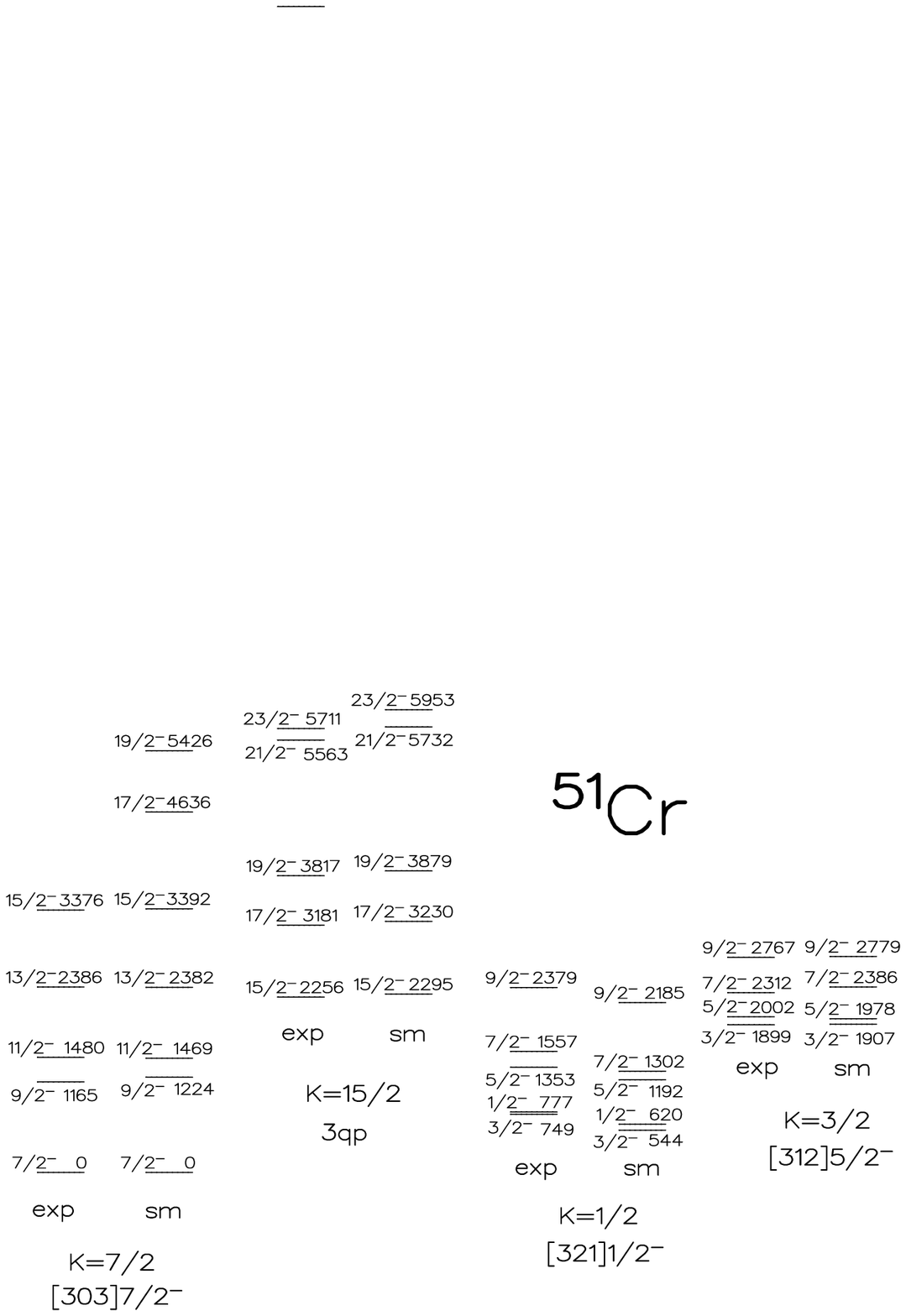,width=8.4cm}
 \protect\caption{Comparison of experimental negative parity levels in $^{51}$Cr with SM predictions ($s$=3). }
 \label{fig07}
 \epsfig{file=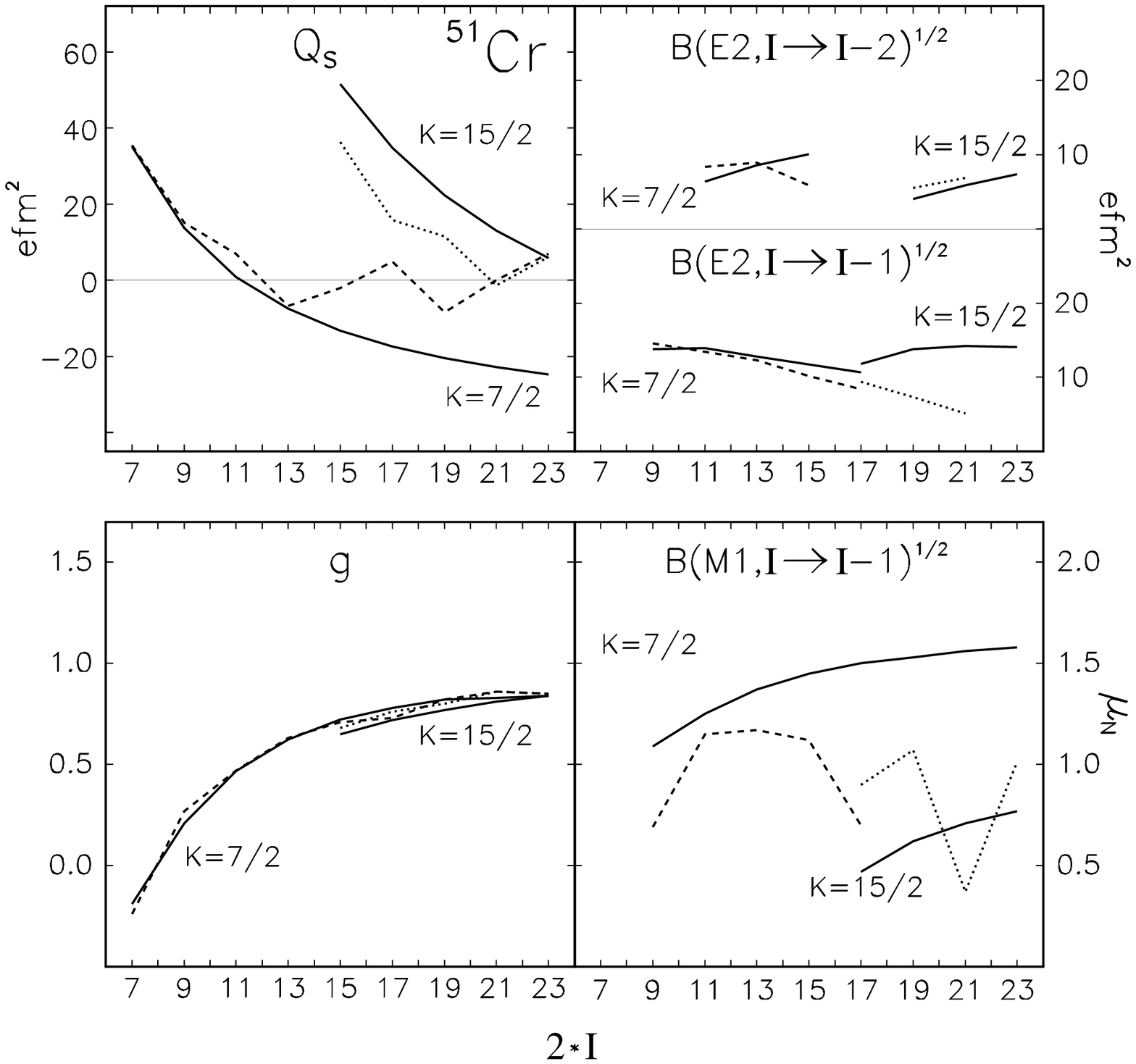,width=8.4cm,angle=0}
 \protect\caption{Calculated em moments in $^{51}$Cr. 
                  Rotor ($\beta^*$=0.21): solid lines, SM: dashed ($K$=5/2) and  dotted ($K$=17/2) lines.}
 \label{fig08}
\end{figure}

\subsection{$^{51}$Cr}

The $N$=$Z$+3  nucleus $^{51}$Cr has been recently discussed \cite{KB3G}, but the bandcrossing  of the gs band was not recognised. The gs band is based in this case on the Nilsson orbital $\nu$[303]7/2$^-$.
 The experimental levels, taken from NDS, are 
displayed in Fig.~\ref{fig07} where they are compared with SM calculations up 
to the termination at 23/2$^-$ ($Q_s$$=$6.9 efm$^2$). 
Since its configuration is $\pi^{-4}\nu^{-1}$, the positive value is due 
to the neutron hole configuration, which is larger than that of the $\nu^{-3}$ 
configuration in $^{49}$Cr. The  calculated levels reported here were obtained with a $s$=3 truncation.
Calculations were made also in full $pf$ configuration space, but a worse average agreement by about $\simeq$ 50 keV was obtained, possibly pointing to a somewhat too large contribution of the $1f_{5/2}$ orbital.
 Relevant SS occurs at the yrast 9/2$^-$ level, but it decreases above, in contrast with $^{49}$Cr. The deformation parameter at low spins is $\beta^*$=0.21, much smaller than in $^{48}$Cr. 
   
The yrast 15/2$^-$ level is interpreted as the head of the 3--qp band obtained by 
lifting one proton from the [321]3/2$^-$ orbital to the [312]5/2$^-$ one and 
coupling the three unpaired nucleons to the maximum value of K. 
The empirical $g$-factor is $g=\mu/I \simeq 2/15[1.38(3/2+ 5/2) -0.30\cdot7/2)]$=0.60, while the SM value is 0.69.

The em observables are plotted in Fig.~\ref{fig08} and they resemble qualitatively the ones of  $^{49}$Cr.
  
\begin{figure}[t]
 \epsfig{file=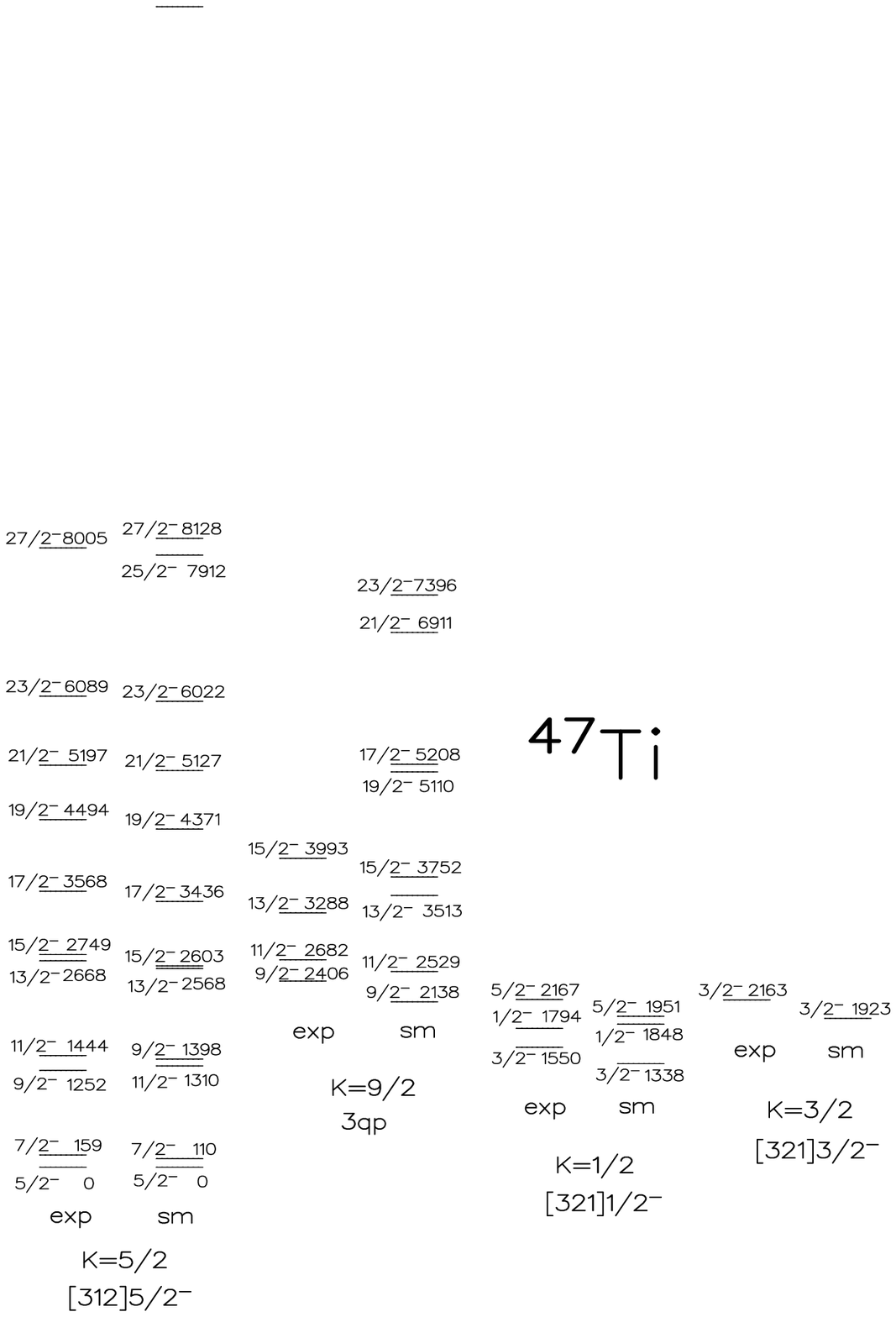,width=8.5cm}
 \protect\caption{Comparison of experimental negative parity levels in $^{47}$Ti with SM predictions. }
 \label{fig09}
 \epsfig{file=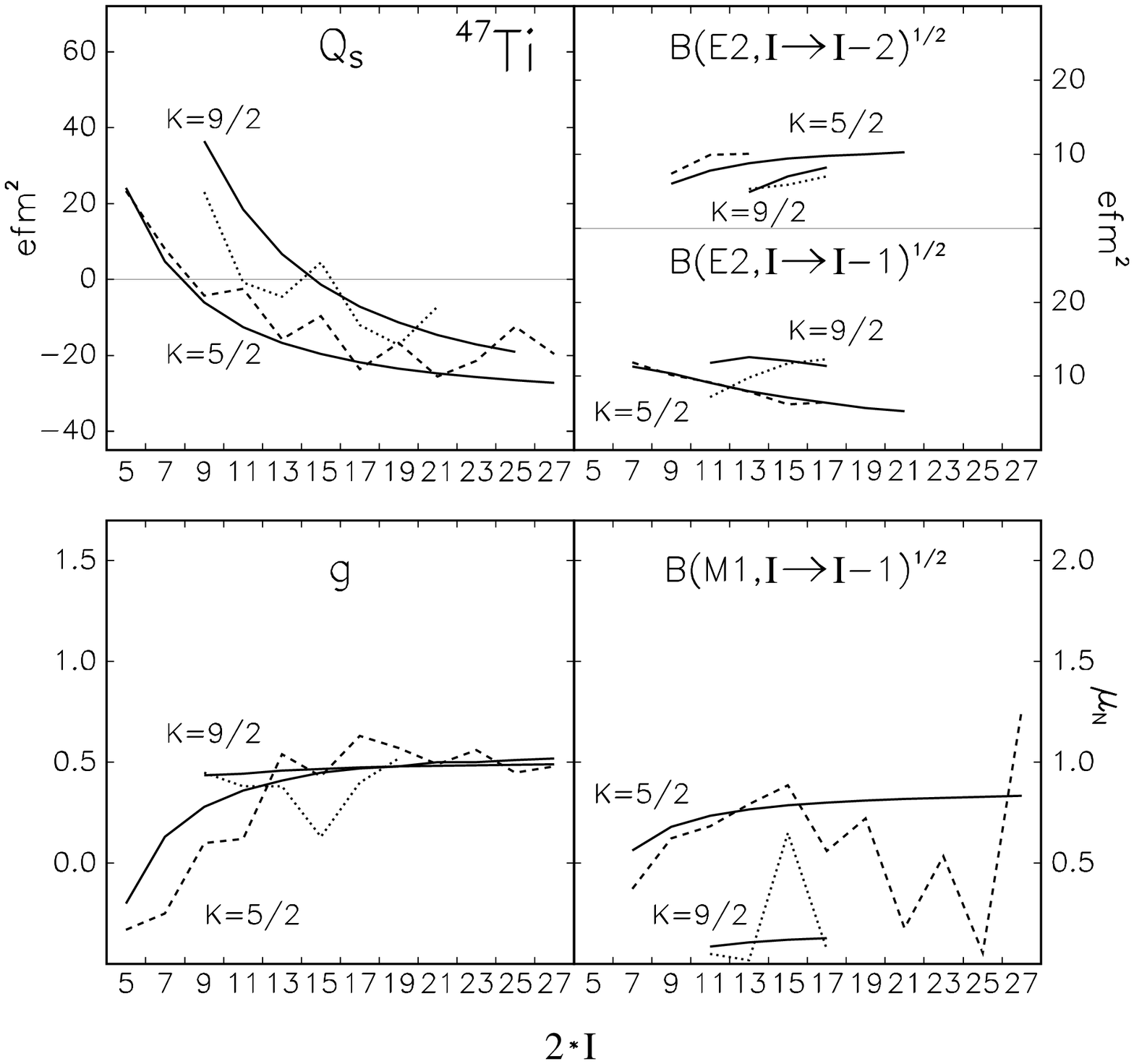,width=8.0cm,angle=0}
 \protect\caption{Calculated em moments in $^{47}$Ti. 
                  Rotor ($\beta^*$=0.22): solid lines, SM: dashed  ($K=$5/2) and dotted  ($K$=17/2) lines. 
		 }
 \label{fig10}
\end{figure}

The $g$-factor values of both bands converge at $I$=23/2 to the empirical value 0.87 (the SM value is 0.85). The agreement of SM values with rotor prediction is remarkable.

\subsection{$^{47}$Ti}

In the $N$=$Z$+3 nucleus $^{47}$Ti the gs band is based on the $\nu[312]5/2^-$ 
orbital as in $^{49}$Cr and shows relevant SS as in $^{51}$Mn. Levels were taken from NDS.  As shown  in Fig.~\ref{fig09} no bandcrossing 
occurs along the gs band, since the $K$=9/2  3-qp band, resulting from the lifting of one proton from the [321]1/2$^-$ orbital to the [312]3/2$^-$ one, is largely non yrast. 
The semiclassical $g$-factor  of the sideband head is $g=\mu/I \simeq 2/9[1.38(1/2+3/2) -0.30\cdot 5/2)]$=0.45 while the SM value is 0.44.
The terminating $Q_s$ value is --19.6 efm$^2$, showing that the shape associated to the $\pi^2$ 
configurations largely prevails on that of the $\nu^{-3}$ one. This occurs 
also in a restricted $f_{7/2}^n$ space.
The empirical $g$-factor value at termination is 0.45, in agreement with the SM value 0.48.

 $^{47}$Ti has been discussed in detail in
Ref.~\cite{Mart2} as a distinct case of incipient rotational collectivity,  
since the $Q_s$ values follow the expectation for a rotating prolate nucleus nearly up to the 
termination at $I$=27/2, but this is illusory since approaching  termination, 
the negative value of $Q_s$ is associated to an oblate non--collective shape and not to a prolate collective one. 
The values of $Q_s$ displayed in Fig.~\ref{fig10} do not reveal the change of regime, which is, however, signalled by the sudden decrease of SS above 15/2$^-$, as in $^{51}$Mn and $^{51}$Cr. The nucleus is not much deformed, so that configurations are quite mixed.
The particularly large perturbation of the sideband at 15/2 can be interpreted as due to the mixing with a 3-$\nu$
configuration. In this case a $K$=15/2$^-$  band results from the parallel coupling of neutron  orbitals [321]3/2, [312]5/2 and [303]7/2$^-$. This produces both the positive value of $Q_s$ and the small value of g. Similarly, the inversion of sideband levels 19/2 and 17/2 may be caused by the interaction with a 19/2 5-qp band, which is obtained by further breaking a neutron pair.
 The  SS at low spins presents a problem since PRM and CSM with $\gamma\le 0$ 
 cannot explain its entire size. This
   may point to some limits of the meanfield approximation.

\subsection{$^{47}$V and $^{49}$V}

Large SS at low spin is observed in the odd V isotopes 
$^{45}$V, $^{47}$V and $^{49}$V. Their level schemes are basically similar at low excitation energies, as 
they are all based on the [312]3/2$^-$ Nilsson orbital, giving rise to a 
$K^\pi$=3/2$^-$ band. The three isotopes are characterized by having very close 
yrast 3/2, 5/2 and 7/2 levels, even if the ordering changes. This can be 
explained by a sensible RAL, induced by the mixing with the orbital 
[330]1/2$^-$ caused by pairing, which is reasonably well reproduced by PRM 
calculations. Termination occurs at 27/2$^-$ in $^{45}$V and $^{49}$V, while 
at 31/2$^-$ in $^{47}$V. The 3--qp band is expected to have K$=$7/2, 11/2 and 
15/2 in $^{45}$V, $^{47}$V and $^{49}$V, respectively.
    
\begin{figure}[t]
\epsfig{file=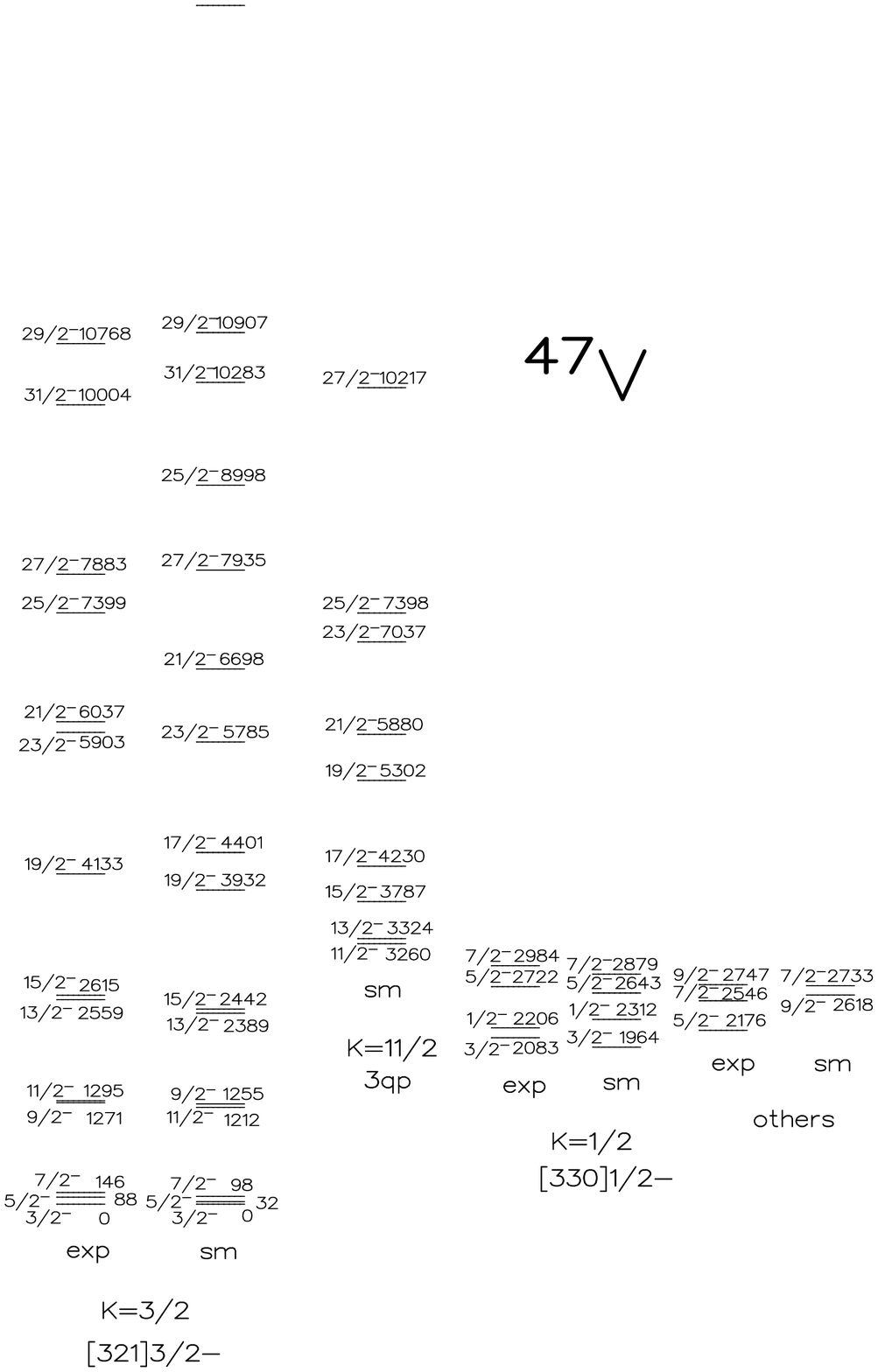,width=8.5cm,height=9.cm,angle=0}
 \protect\caption{Comparison of experimental negative parity levels in $^{47}$V with SM predictions.}
 \label{fig11}
 \epsfig{file=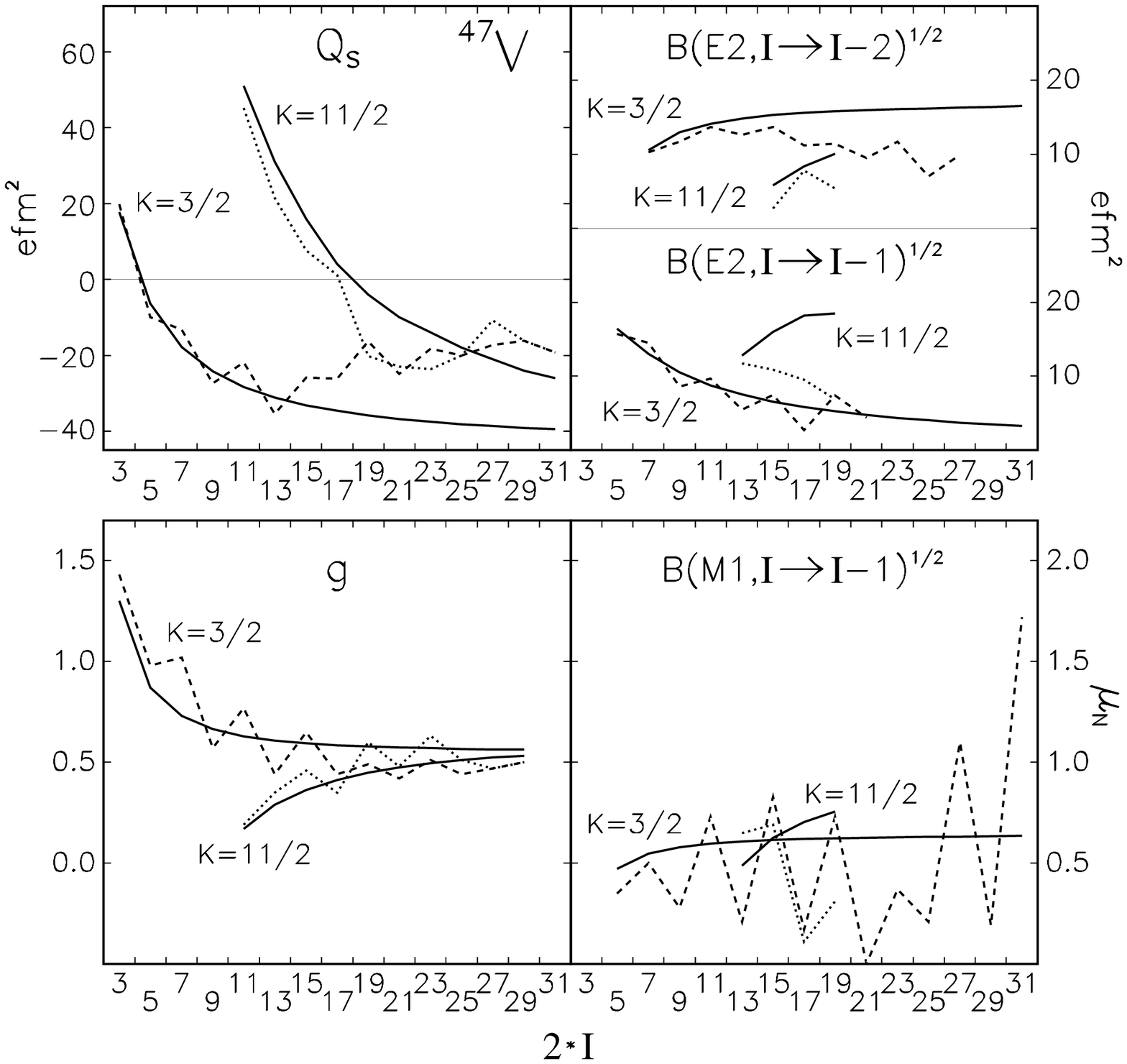,width=8.4cm,angle=0}
 \protect\caption{Calculated em moments in $^{47}$V. Rotor ($\beta^*$=0.26): solid lines,   SM: dashed  ($K$=5/2) and dotted  ($K$=11/2) lines. }
 \label{fig12}
\end{figure}

\begin{figure}[t]
 \epsfig{file=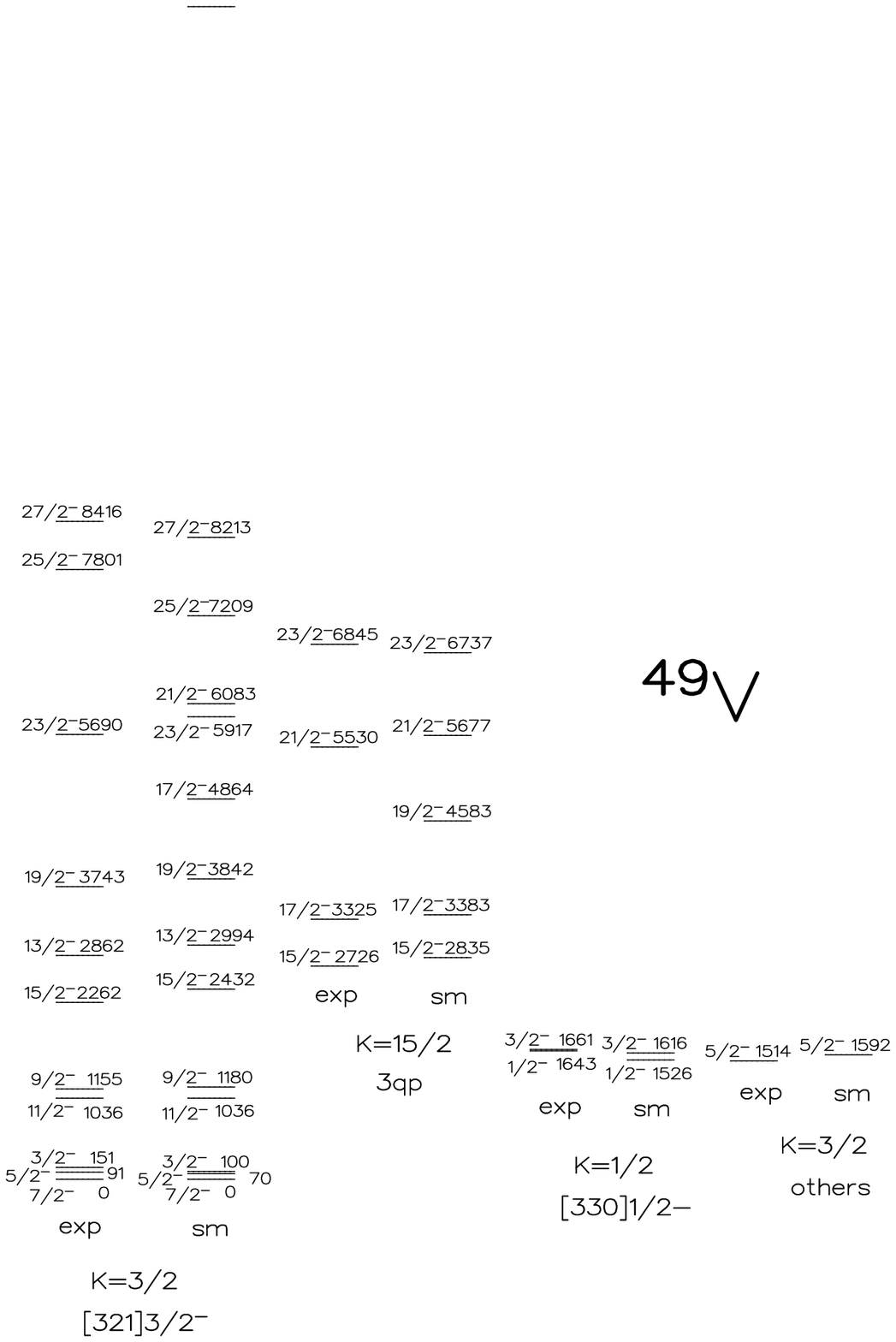,width=8.5cm,}
 \protect\caption{Comparison of experimental negative parity levels in $^{49}$V with SM predictions.}
 \label{fig13}
 \epsfig{file=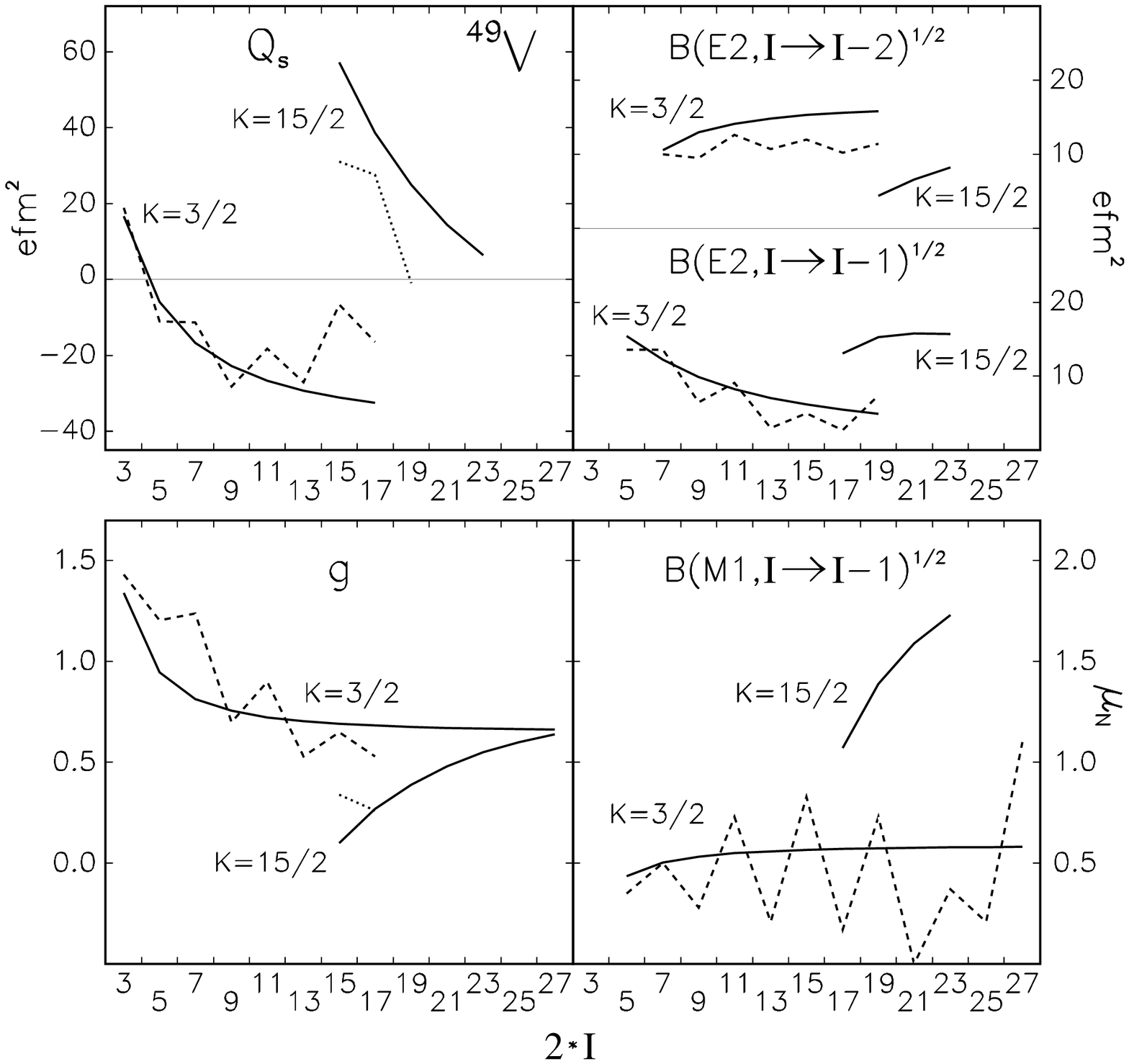,width=8.5cm,angle=0}
 \protect\caption{Calculated em moments in $^{49}$V. Rotor ($\beta^*$=0.25): solid lines,  SM: dashed   ($K$=5/2) and  dotted  ($K$=15/2) lines. }
 \label{fig14}
\end{figure}

The level scheme of the $N$=$Z$+1 nucleus $^{47}$V is shown in Fig.~\ref{fig11}. The gs band has been discussed in Ref.~\cite{Bra47V49CR}, where the
experimental $B(E2)$ and $B(M1)$ values along the gs band turned out to agree reasonably with theory. Its deformation at low spins is nearly as large as that of $^{49}$Cr, but  mixing is larger than in $^{49}$Cr.
The calculated 17/2$^-$ level belonging to the gs band 
is the yrare one, while the yrast 17/2$^-$ level belongs essentially to the 
predicted K$=$11/2 3-qp band~\cite{BraGASP}. Unfortunately, both 17/2$^-$ levels were not observed experimentally. The empirical $g$-factor of the $K=11/2^-$ band head 
is $g=\mu/I \simeq 2/11[1.38\cdot3/2 -0.30(3/2+5/2)]$=0.16 to be compared with the SM value 0.18. According to Fig.\ref{fig11} also the yrast levels 21/2$^-$ and 
25/2$^-$ belong essentially to the sideband, even if they are mixed.
 As shown in Fig.~\ref{fig12}, staggering is observed at low spin for both $Q_s$ and g. 
This may be associated with triaxiality but it is not easy to disentangle 
the effects of CD. 

The $Q_s$ values above $I^\pi=17/2^-$ become 
smaller  and remain constant up to the termination at 31/2$^-$ 
($Q_s=-14.4$ efm$^2$). 
The empirical $g$-factor value at the termination is 0.51, in agreement with the the SM value 0.53.
The SS increases with spin leading to ordering 
inversion. According to Ref.~\cite{Mart2} the $g$--factor values around 19/2$^-$ 
reflect a  $v$=3 termination but, contrarily to $^{49}$Cr, the change of 
regime seems to interfer with the effects of $v=$3 termination in such a 
way  that the backbending at $I$=19/2 is not observed.
 In this nucleus it is particularly difficult to assign side levels to 1-qp bands, since there is no regularity. The $K$=1/2 band is denoted by
[330]1/2$^-$. The experimental 5/2 level at 2176 does not have a calculated counterpart. 
        
 Fig.~\ref{fig13} shows that the level scheme of the $N$=$Z$+3 nucleus $^{49}$V is well predicted by SM calculations and it presents  analogies with that of $^{47}$V.  At the termination the $Q_s$ value is --8.5 efm$^2$, showing that the proton hole shape prevails. It has the largest SS among the three isotopes and its gs band is expected 
to be crossed by a $K^\pi$=15/2$^-$ 3-qp band, which is observed experimentally. 
Its bandhead is the yrare 15/2$^-$ with $Q_s=33.0$ efm$^2$, while the member 
with $I$=17/2 is yrast. 
The larger SS is probably related with triaxiality, reflecting 
a high configuration mixing among Nilsson configurations. 
 For this reason a limited number of levels is reported in  Fig.~\ref{fig14}.  For sidebands only theoretical predictions are displayed in this case.
The $g$-factor of the $K$=15/2 band head is predicted to be 
$g=\mu/I \simeq 2/15[1.38\cdot3/2 -0.30(5/2+7/2)]$=0.04, which greatly differs from the SM value of 0.34, indicating mixing effects, which apparently also lower its $Q_s$ value in Fig.~\ref{fig14}.

The empirical $g$-factor value at the termination is 0.63, in agreement with the the SM value 0.66.

    
The level schemes of $^{45}$V and of its mirror $^{45}$Ti are identical, apart for 
Coulomb effects. Their level schemes are somewhat perturbed at low spin by 
intruder 2h-- and 4h--configurations. They will  not be further discussed here because  their level schemes are not sufficiently well known to bring relevant new information.

\subsection{$^{48}$Cr}
 
\begin{figure}[t]
\vskip-0.5cm
 \epsfig{file=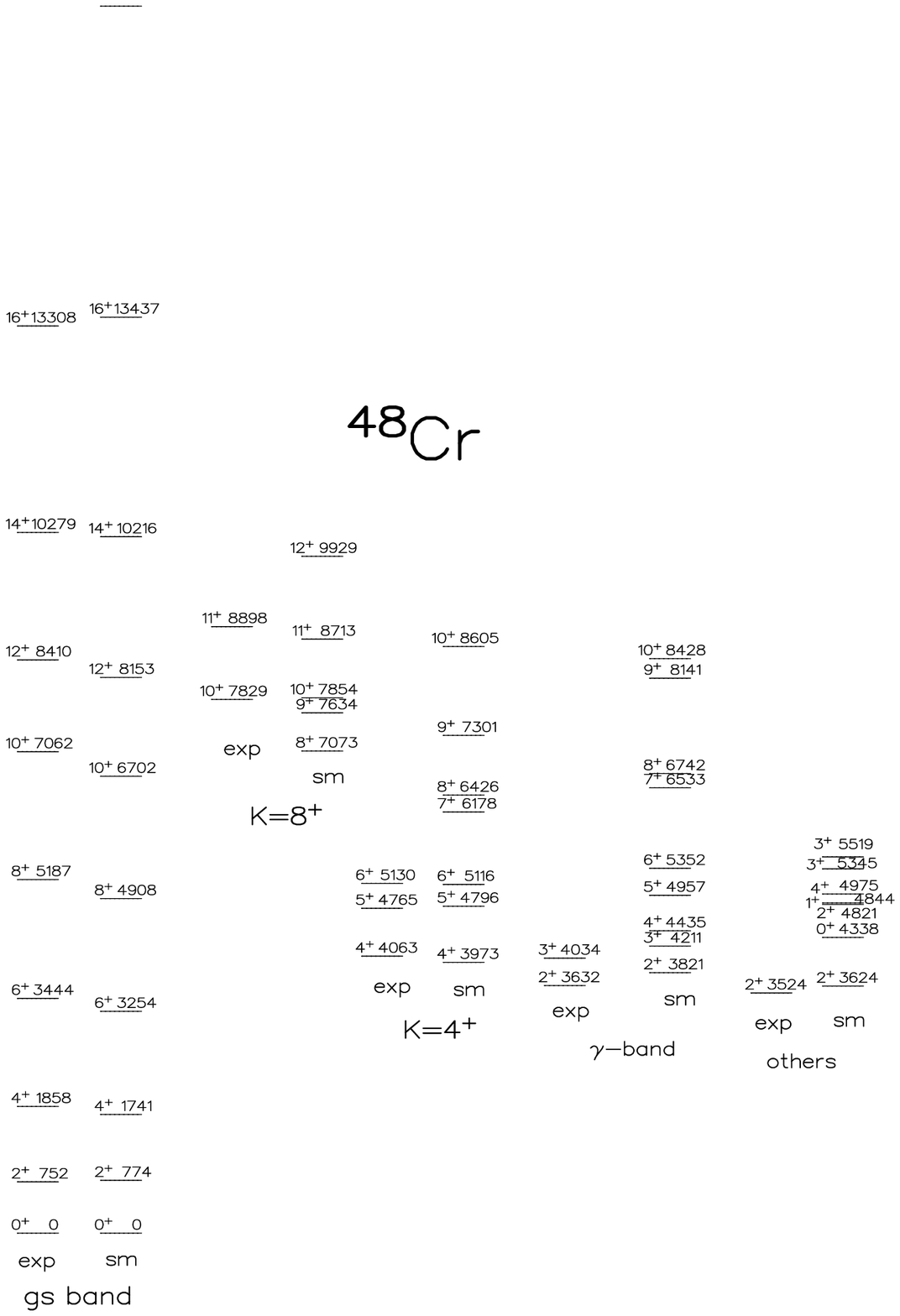,width=8.5cm,height=9.5cm}
 \protect\caption{Comparison of experimental positive parity levels in $^{48}$Cr with SM predictions. }
 \label{fig15}
 \epsfig{file=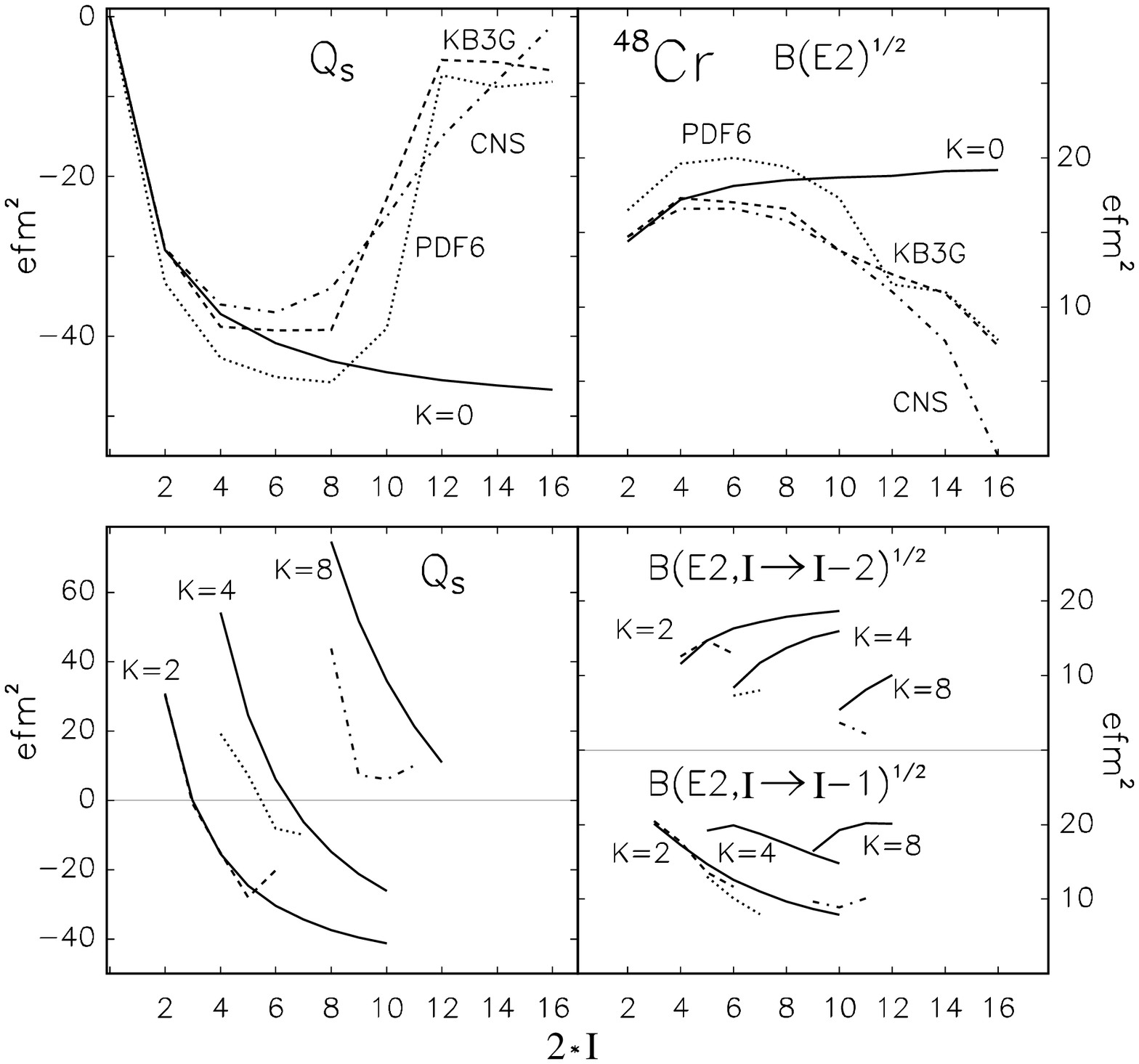,width=8.4cm}
 \protect\caption{Upper panels) Comparison of rotor ($\beta^*$=0.31) estimates with some SM   predictions for the gs band  in   $^{48}$Cr. Lower panels) Calculated em moments in $^{48}$Cr. Rotor: solid lines,  SM: dashed   ($K$=2),  dotted  ($K$=4) and dot-dashed ($K$=8) lines.}
 \label{fig16}
\end{figure}

In the last ten years $^{48}$Cr has been addressed by several theoretical 
studies as its understanding is considered to be of strategic importance.
An updated experimental level scheme is reported in Fig.~\ref{fig07}. Yrast 
levels were taken from Ref.~\cite{Bra48CR50}, while yrare levels essentially 
from a recent work~\cite{Jess}, where spin--parity assignments were not 
precisely determined. It is assumed that all experimental sidelevels in Fig.~\ref{fig15}
have positive parity, on the basis of the following argument: in $^{48}$Cr 
a negative parity band with $K^\pi$=4$^-$, whose band head is at 3532 keV, was 
determined in Ref.~\cite{Bra48CR50}, where it was  described 
as the coupling of the [202]3/2$^+$ orbital with the [312]5/2$^-$ one.
This band was well reproduced by SM calculations, which predict at an energy more than one MeV higher the $K^\pi$=1$^-$ band originating from the antiparallel coupling.  The reliability of these predictions is stressed by the fact that the lowest terms of 
both bands were observed and correctly reproduced in $^{50}$Cr~\cite{BraGASP}.

The observed lowest sidelevels \cite{Jess} are tentatively assigned to the calculated bands according to their energy and decay scheme. The very good agreement cannot be considered fortuitous.  The original SM calculations using KB3 \cite{Caur2} differ from the present ones by  at most hundred keV for any reported level and predict very similar em moments within few percents. In the rest of the subsection we will refer principally to calculated level energies.
The structure of the $^{48}$Cr gs band was extensively discussed in the frame of
SM~\cite{Caur2}. As shown in Fig.~\ref{fig15} the gs band develops without  
band crossing up to the termination at 16$^+$. This was also confirmed  by 
CHFB calculations~\cite{Caur2,Tanaka}. In Fig.~\ref{fig16} the deformation parameter $\beta^*$=0.31 is assumed for the initial deformation, as previously made (Ref.~\cite{Bra48CR50}).
The lowest 2--qp band is formed with $K^\pi$=4$^+$ by promoting one neutron or one 
proton from the [321]3/2$^-$ to the [312]5/2$^-$ orbital~\cite{BraSev}. Its head is the yrare 4$^+$ which lies 2.3 MeV above the yrast 4$^+$. It is rather unexpected that its $Q_s$ value is less than half the rotor prediction (Fig.~\ref{fig16}, lower panels), but anyhow the band develops rather regularly up to 8$^+$.
The predicted band shows  SS and the $B(E2)$ to the fourth 4$^+$ level at 4975, built also mainly on a $1f_{7/2}^n$ configuration,  is 163.0 efm$^2$.
  These feature may indicate  triaxiality which could explain the low $Q_s$ value of the yrare 4$^+$ level. The yrare 2$^+$ level at 3624 keV has  $Q_s$=20.7 efm$^2$, consistent with a $K$=2 bandhead with a reduced deformation and it is connected with a $B(E2)$ rate of 104.0 efm$^2$ to the yrare $3^+$ level at 5345 keV.
 
 The antiparallel coupling gives rise to a 2-qp $K^\pi$=1$^+$ band. The lowest 1$^+$ level is predicted at 4844 keV. The corresponding experimental level is not known, but the prediction has to be considered reliable since in $^{50}$Cr 
the yrast 1$^+$ is at 3629 keV, while theory gives the precise estimate of 3539 keV.
A  favoured connection of the yrast 1$^+$ level with the fourth 2$^+$ level at 4821 keV ($B(E2)$=280.6 efm$^2$) is the only
indication of an incipient $K^\pi$=1$^+$ band, which gets exhausted  with the third 3$^+$ level at 5519 keV ($B(E2)$=86.1 efm$^2$). 
All such calculated levels, without clear collective properties, are grouped on 
the rightmost side of Fig.~\ref{fig15}. 
         
As shown in the same figure, a 4--qp band with $K^\pi$=8$^+$ is predicted to start  
at an  excitation energy of 7073 keV, about twice larger  that for the $K^\pi$=4$^+$ band, being due  to the simultaneous breaking of a proton and a  neutron pair.  Its $Q_s$ value 
is 43.8 efm$^2$, corresponding to $\beta^*$=0.18. The observed yrare 
10$^+$ level  decays mainly to the yrast 8$^+$ level with a 2644 keV transition and has a weak branch of 768 keV to the yrast  10$^+$ one~\cite{Ur2}. This is expected  by SM calculations for the member of  the $K^\pi$=8$^+$ band.
 The yrare 12$^+$ is predicted to lie 2 MeV above the yrast one and to belong essentialy to the 
$K^\pi$=8$^+$ band. At a similar excitation energy of the $K^\pi$=8$^+$ band also 
a 4--qp $K^\pi$=2$^+$ band may be expected of which however no evidence exists.
       
The $\gamma$--band is predicted to start at 3821 
keV and to be well deformed (Fig.~\ref{fig16}, lower panels). The peculiar nature of this  
band is revealed by its very fragmented wavefunction, where the 1$f_{7/2}^n$ 
configuration contributes to less than 1\% to the wavefunction of the lowest members, while in the gs band it has a 
20\% contribution and in the $K^\pi$=4$^+$ band a 37\% one. Above 6$^+$ relevant mixing with the $K=4$ band occurs and also a large SS.

The $Q_s$ values of the gs band approaching termination are also very peculiar (Fig.~\ref{fig16}, upper panels): they are 
consistent with an axially prolate description up to $I^\pi=8^+$ but then 
 suddenly become nearly zero at backbender level 12$^+$ and remain so for the levels  14$^+$ and 16$^+$.
If one repeats the calculations with the FPD6 interaction the behaviour is 
even more pronounced.
 CHFB~\cite{Caur2,Tanaka} and CNS~\cite{Juo2} calculations estimate 
$\gamma\simeq -15^\circ$ at the backbending region. In Fig.~\ref{fig16} the 
CNS predictions are displayed for comparison. The axial description is good 
up to 6$^+$, while the	sudden decrease of $Q_s$ above it requires some  
triaxiality with $\gamma\le 0$. In fact, in the case of rotational collectivity $Q_s$ is related to $Q_\circ$ by eq.~(\ref{eq01}), which for $K$=0 becomes $Q_s=-Q_\circ I/(2I+3)$  where:
		   
\begin{equation}
 Q_\circ=\frac{6}{\sqrt{5 \pi}}Z e R^2_0 \beta \sin(30 +\gamma)
 \label{eq07}
\end{equation}
	    
For  negative values of $\gamma$ $Q_\circ$ decreases with respect to the axial symmetry, while an increase of $B(E2)$ values should occurs, according to eq.~(\ref{eq02}) and the relation:

\begin{equation}
 Q_t=2 \sqrt{\frac{3}{5 \pi}}Z e R^2_0 \beta \cos(30 +\gamma)
 \label{eq08}
\end{equation}
 
One should, however, comment that neither CNS~\cite{Juo2} nor standard CHFB 
calculations~\cite{Caur2} can reproduce well the  backbending and 
the drastic reduction of $Q_s$ at $I^\pi=12^+$, which may suggest the 
bandcrossing with the 4--qp  $K^\pi$=2$^+$ band, claimed in a PSM analysis~\cite{Ha}.
 However, the presence  of that band, described in the PSM as little deformed,  is excluded   by SM and experimental data, as previously discussed.
Since the yrare 12$^+$ level lies about 2 MeV higher than the yrast one, it 
has, on the contrary, to be concluded that what occurs is an intrinsic change 
of structure, not well described in a mean field approximation.
In analogy to $^{49}$Cr, the discontinuity at 12$^+$ can be interpreted as a 
sudden transition from collective to a nearly spherical shape, persisting at 
spins 14 and 16 and probably related with the $v$=4 termination at 12$^+$, in absence of bandcrossing.
                 
\subsection{$^{52}$Fe and $^{44}$Ti}

\begin{figure}[t]
 \epsfig{file=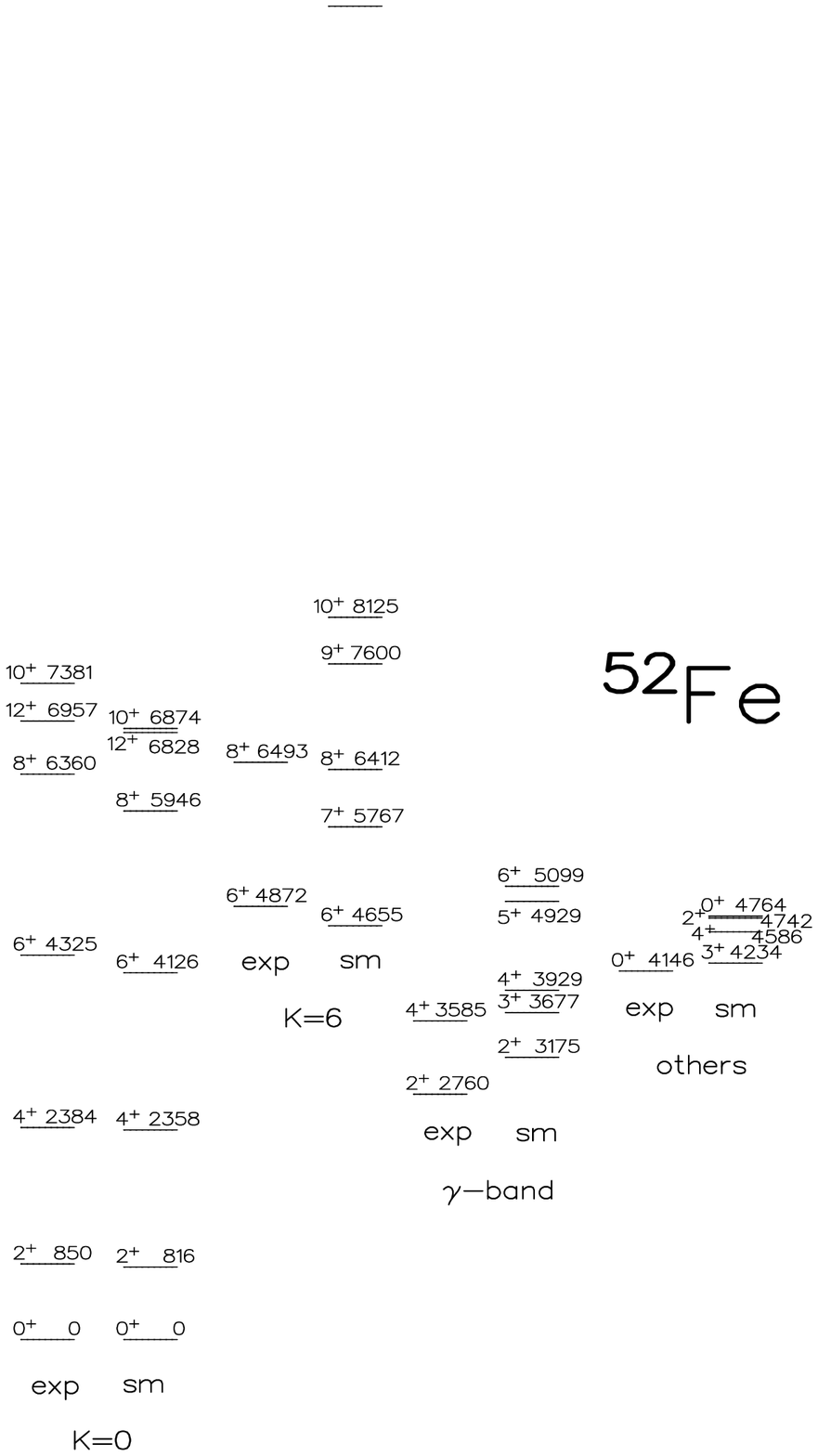,width=8.5cm}
 \protect\caption{Comparison of experimental positive parity levels in $^{52}$Fe with SM predictions. ($s$=3)}
 \label{fig17}
 \epsfig{file=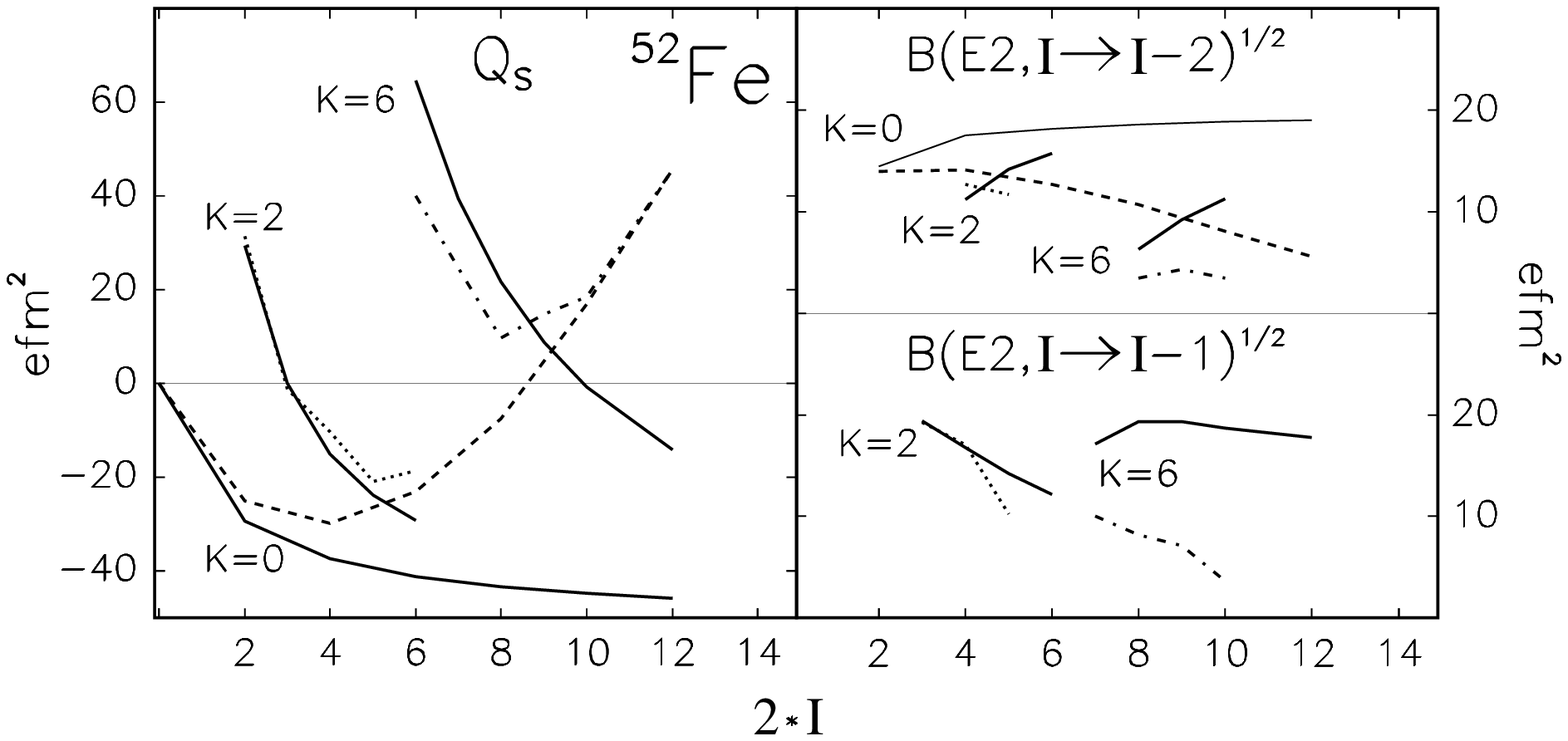,width=8.cm,angle=0}
 \protect\caption{ Calculated em moments in $^{52}$Fe. Rotor ($\beta^*$=0.26): solid lines,  SM: dashed   ($K$=0),  dotted  ($K$=2) and dot-dashed ($K$=6) lines. See text for details.}
 \label{fig18}
\end{figure}

The experimental level scheme of $^{52}$Fe is taken from NDS and from Ref.~\cite{Ur}. 
 The fact that the terminating state 12$^+$ 
lies 424 keV below the 10$^+$ has a similar explanation as in $^{53}$Fe.
 As shown in Fig.~\ref{fig17}, the gs band 
in $^{52}$Fe develops rather regularly up to spin 10$^+$, with a smaller 
moment of inertia than $^{48}$Cr.  The calculated scheme  was obtained with $s$=3. Clear correspondence is evident between experiment and theory, despite  the somewhat worse  agreement than for $^{48}$Cr.
 
\begin{figure}[t]
 \epsfig{file=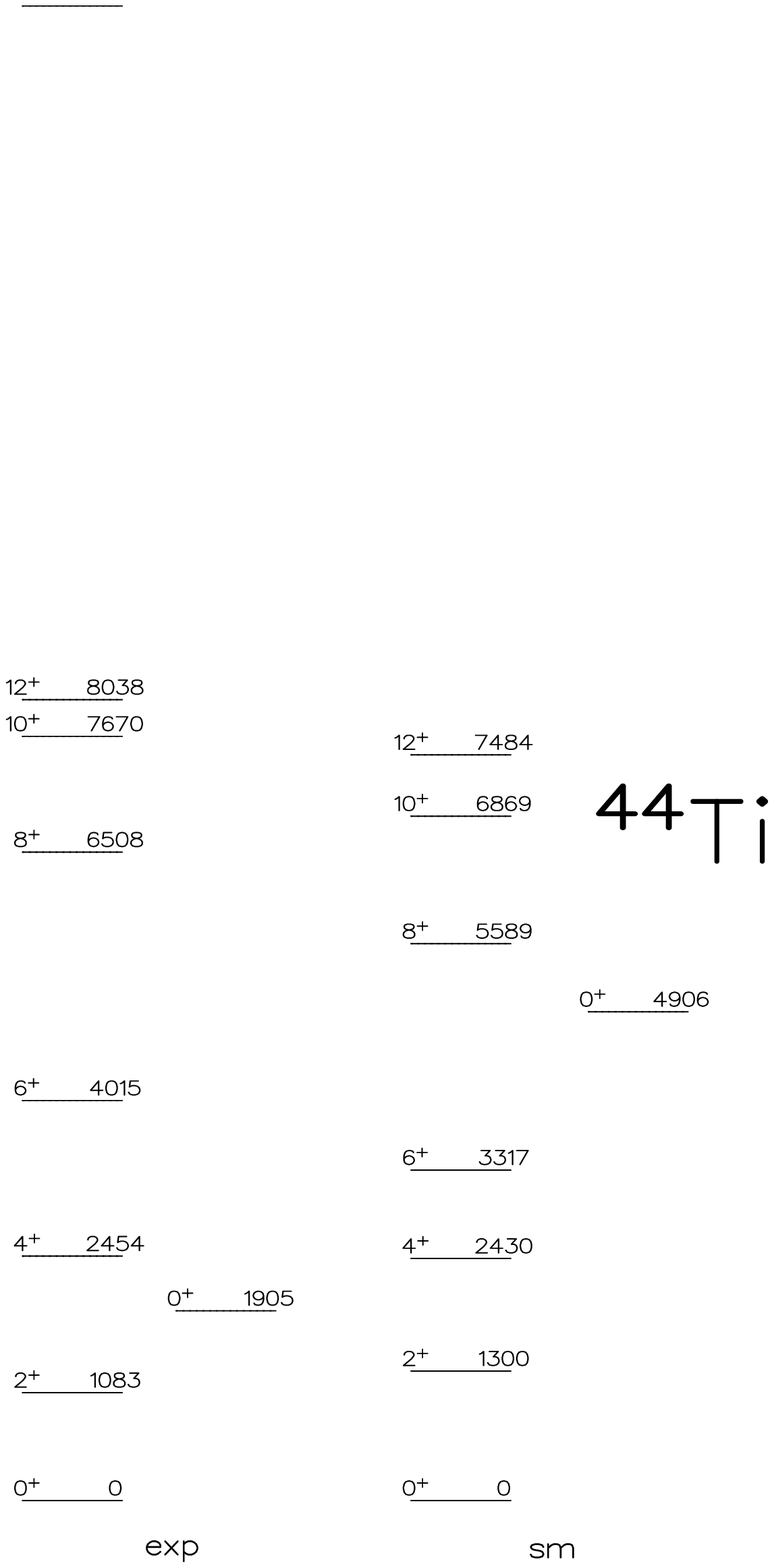,width=6.cm,angle=0}
 \protect\caption{ Comparison of experimental gs band and yrare 0$^+$ levels in $^{44}$Ti with SM predictions. }
 \label{fig19}
 \epsfig{file=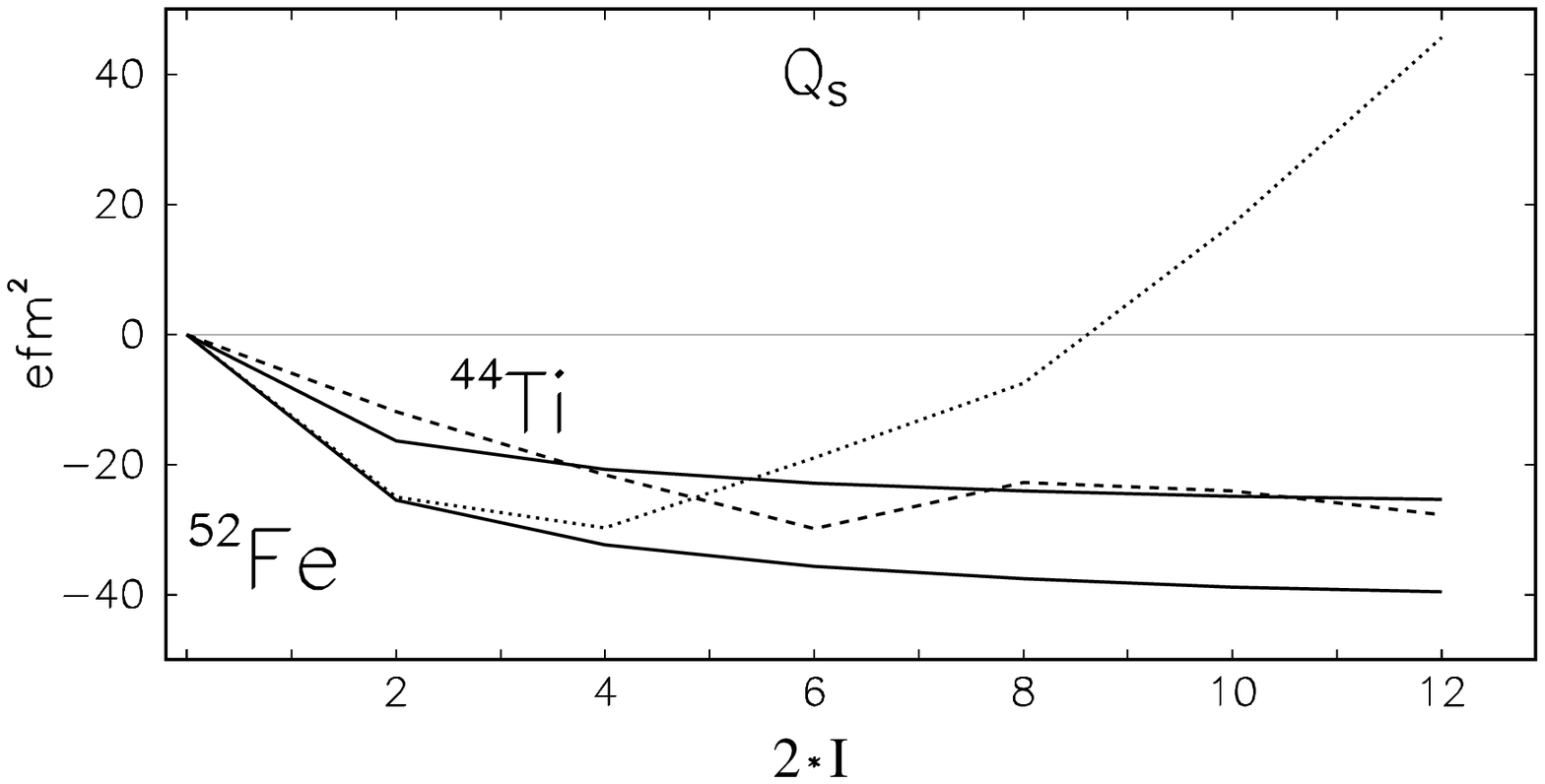,width=8.cm,angle=0}
 \protect\caption{Comparison between experimental and
                  theoretical gs band  $Q_s$ values in $^{52}$Fe ($\beta^*$=0.23) and $^{44}$Ti ($\beta^*$=0.21).}
 \label{fig20}
\end{figure}

The presence of two close yrare levels 6$^+$ and 8$^+$    
 was already related to the  2--qp K$=$6 band, formed 
by the excitation of one neutron or one proton from the [312]5/2$^-$ to the 
[303]7/2$^-$ orbital~\cite{Ur}. The experimental 
yrare 2$^+$ and 4$^+$ levels have, on the other hand,  to be attributed to the 
$K^\pi$=2$^+$ $\gamma$--band, which lies about one MeV lower than in $^{48}$Cr, 
in spite of the lower deformation. As in $^{48}$Cr, the composition 
of $\gamma$--band is very fragmented, while in the $K^\pi$=0$^+$ and $K^\pi$=6$^+$  bands the $(f_{7/2})^{-4}$ configuration is dominant.
The relevant offset for the $\gamma$--band with respect to calculations can, at least partially, be attributed to the space truncation. In fact it has been empirically observed 
that the excitation energies of the $\gamma $-band decrease remarkably  with increasing $s$. The $\gamma$--band corresponds to a deformation $\beta^*$=0.26, as shown in Fig.\ref{fig18}, and develops regularly up to $I$=6. There is, however, a  mixing of about 15\% with the gs band, as the yrare 4$^+$ level is predicted to be connected to the yrast 2$^+$ one by a $B(E2)$ rate of 40 efm$^2$. For this reason one expects that the levels decays mainly to the yrast 2$^+$ level with a 2735 keV transition, while only by about  one percent to the yrare 2$^+$ with a 825 keV transition.


 Calculations predict the inversion between the yrast 12$^+$ and 10$^+$ levels, but their energy spacing is better reproduced by the FPD6 interaction, which, however, often  predicts too large electric moments. Unfortunately $B(E2)$ values are not known with a sufficient precision in $^{52}$Fe to discriminate between different effective interactions but it was shown in subsection A that the KB3G interaction gives a better agreement in the case of $^{54}$Fe. 
It turns out that the KB3G interaction underestimate the relative binding energy of the terminating state in $^{52}$Fe. The same occurs also for the 27/2$^-$ terminating state in $^{51}$Mn.  It seems
  that the interaction KB3G needs an adjustment approaching the shell closure.  The GXPF1 interaction gives  similar results to the  KB3G one.

The $Q_s$ value of the 12$^+$ terminating state is calculated to be 
$47.1$ efm$^2$
 in agreement  with the expected prolate non--collective shape associated to a 4--hole configuration and corresponding to a deformation  $\beta^*$=0.12.
 Strong mixing of the gs band 6$^+$ and 8$^+$ levels with the corresponding  
members of the $K^\pi$=6$^+$ band is indicated by the experimental lifetimes and the connecting 
transitions~\cite{Ur}. Since  large mixing makes the interpretation
difficult, the $Q_s$ data of the K=0 and 6 bands represented in Fig.~\ref{fig18} were obtained in the 
$1f_{7/2}2p_{3/2}$ subspace ($s$=0), where the bands do not interfer much and the 
plots become similar to those in $^{51}$Mn. The value of the terminating $Q_s$ is slightly reduced  to 46.8 efm$^2$, while the  $\beta^*$ parameter decreases by about 8 \% for the yrast 2$^+$ and 4$^+$ levels.
 

Using the Nilsson diagram for predicting a 4--qp band with the simultaneous 
excitation of both proton and neutron pairs, one gets $K=12^+$, which is 
however the terminating state so that there is not a 4--qp band.

In Fig.~\ref{fig19} the experimental levels of the gs band of the cross--conjugate 
nucleus $^{44}$Ti are compared with SM values. The agreement is poor. If one lifts up the SM spectrum by nearly one MeV in order to get  the two 8$^+$ levels at the same height,   the 0$^+$, 2$^+$ and
4$^+$ levels appear to be  more bounded than predicted by about one MeV.
 This is attributed to the strong influence of 2- and 4-hole configurations, as qualitatively discussed in Ref. \cite{OL}, but quantitatively the question is still open. This interpretation is confirmed by the fact that the yrare 0$^+$ is predicted at 4906 keV, while it is observed at 1905 keV. The experimental one is clearly  an intruder state which originates from 2- and 4-hole excitations.
In this view, the levels 6$^+$ should be relative unaffected. 
  Above $I$=6, which is the $v$=2 termination,  the level spacing increases i.e. the opposite of a backbending seems to occur, 
 which may be related with the onset of a change of regime.
  Looking back to $^{48}$Cr one may wonder whether a small effect of the same type  may explain the systematically lower level energies along the gs band  at intermediate spins.

It is worthy to consider  what would happen in absence of such intruder  contributions.
  The $Q_s$ behaviour of $^{52}$Fe is compared with the one of the 
cross--conjugate nucleus $^{44}$Ti in Fig.~\ref{fig20}. The $Q_s$ values do not mark the 
change of regime in $^{44}$Ti, since they are anyhow negative as in $^{47}$Ti. In this case the deformation parameter is reduced in $^{52}$Fe in order to get a fit for the lower spins.
 

\subsection{$^{50}$Cr and $^{46}$Ti}

\begin{figure}[t]
 \epsfig{file=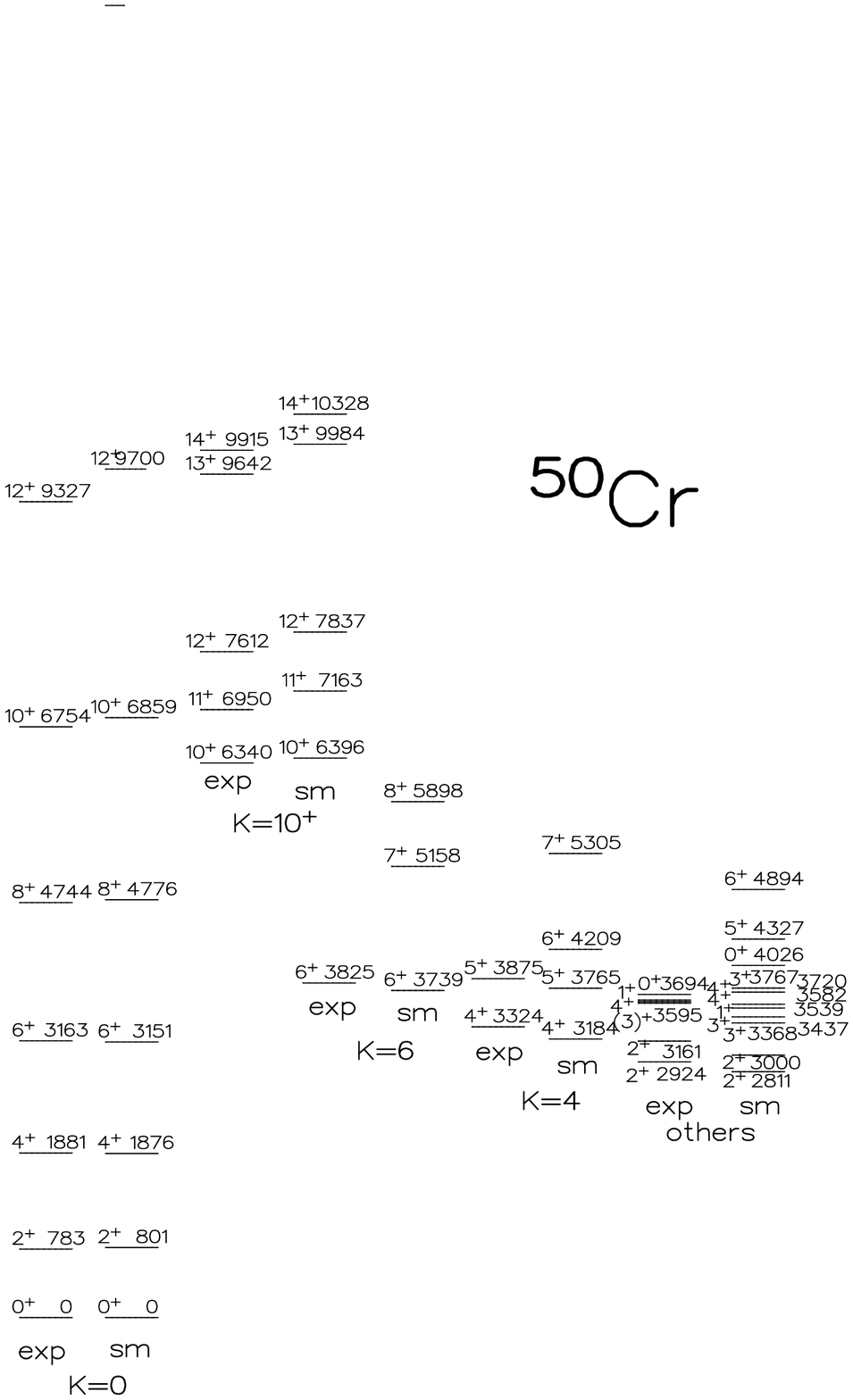,width=8.5cm}
 \protect\caption{Comparison of experimental positive parity and yrare 0$^+$ levels in $^{50}$Cr with SM predictions.}
 \label{fig21}
 \epsfig{file=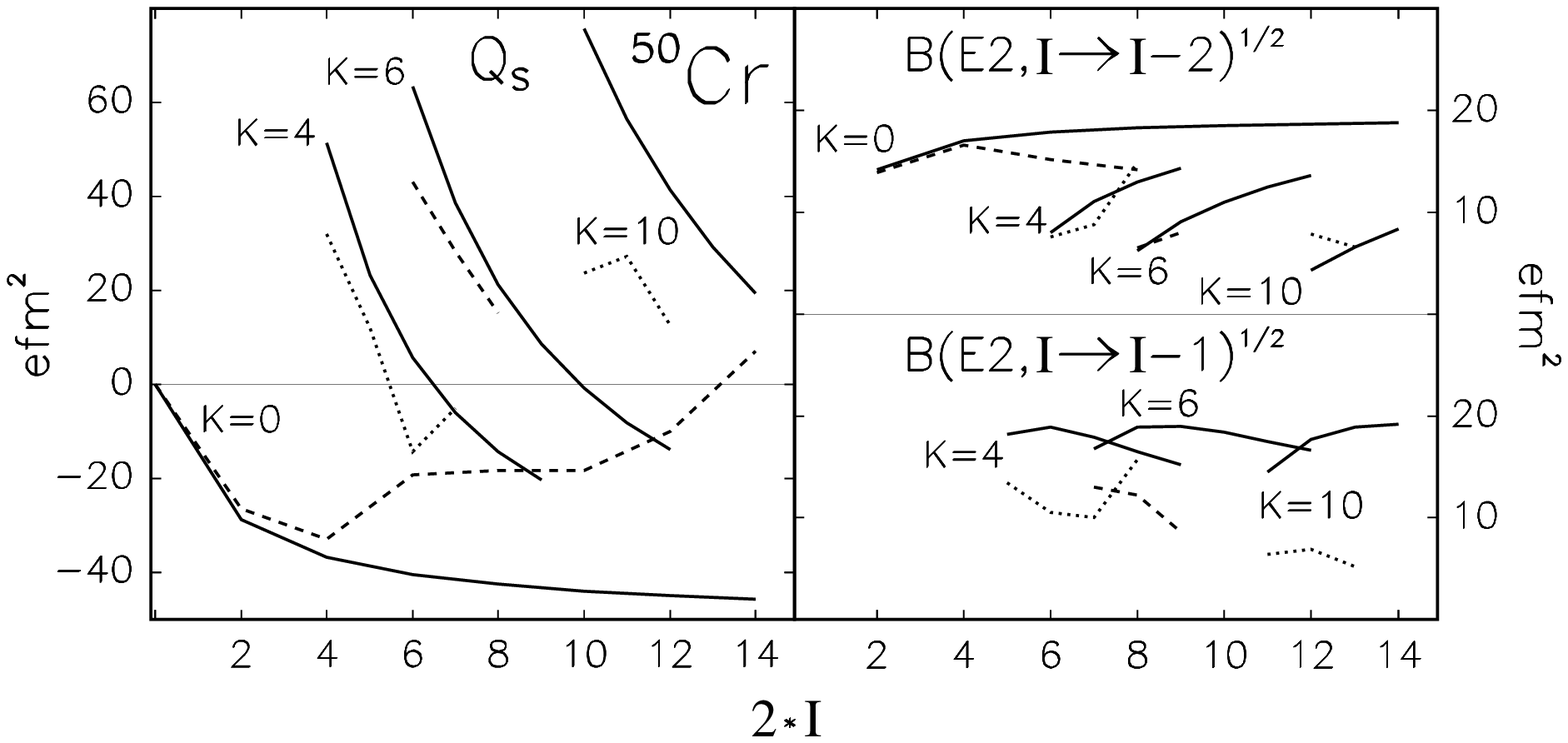,width=8.4cm,angle=0}
 \protect\caption{Comparison between experimental and theoretical  $Q_s$ in $^{50}$Cr. Rotor ($\beta^*$=0.28): solid lines,  SM: dashed   ($K$=0 and 6) and  dotted  ($K$=4 and 10) lines. }
 \label{fig22}
\end{figure}

The $N$=$Z$+2 nucleus $^{50}$Cr has been recently studied in detail~\cite{Bra50CR}. 
The level scheme in Fig.~\ref{fig21} is essentially taken from that reference, even if some more calculated levels are now reported.
 The gs $Q_s$ values are negative at low spin, being related to a collective prolate 
shape, while they become positive  at the terminating level 14$^+$  (7.1 efm$^2$) as expected for an aligned--hole configuration. Bands with $K^\pi$=4$^+$ and $K^\pi$=6$^+$ were 
 observed, which are expected by exciting one proton and one neutron  from the 
[321]3/2 and [312]5/2 respectively. 
 A $K$=10 band was also observed, which is obtained by breaking simultaneously both the neutron and the proton pairs. The em properties are displayed in Fig.~\ref{fig22}. The smaller value $Q_s$ value of the 6$^+$ level is presumably due to the mixing with the head of the $K$=6 band.
This may explain the large $B(M1)$ rate  between the two levels~\cite{Bra50CR}.
The yrast 10$^+$ level is experimentally identified as the head of a $K$=10 band.
SM calculations with the KB3G interaction predict a nearly  50 \% mixing with the 10$^+$ level of the gs band, which has be  commented to be an overestimate~\cite{Bra50CR}.

\begin{figure}[t]
 \epsfig{file=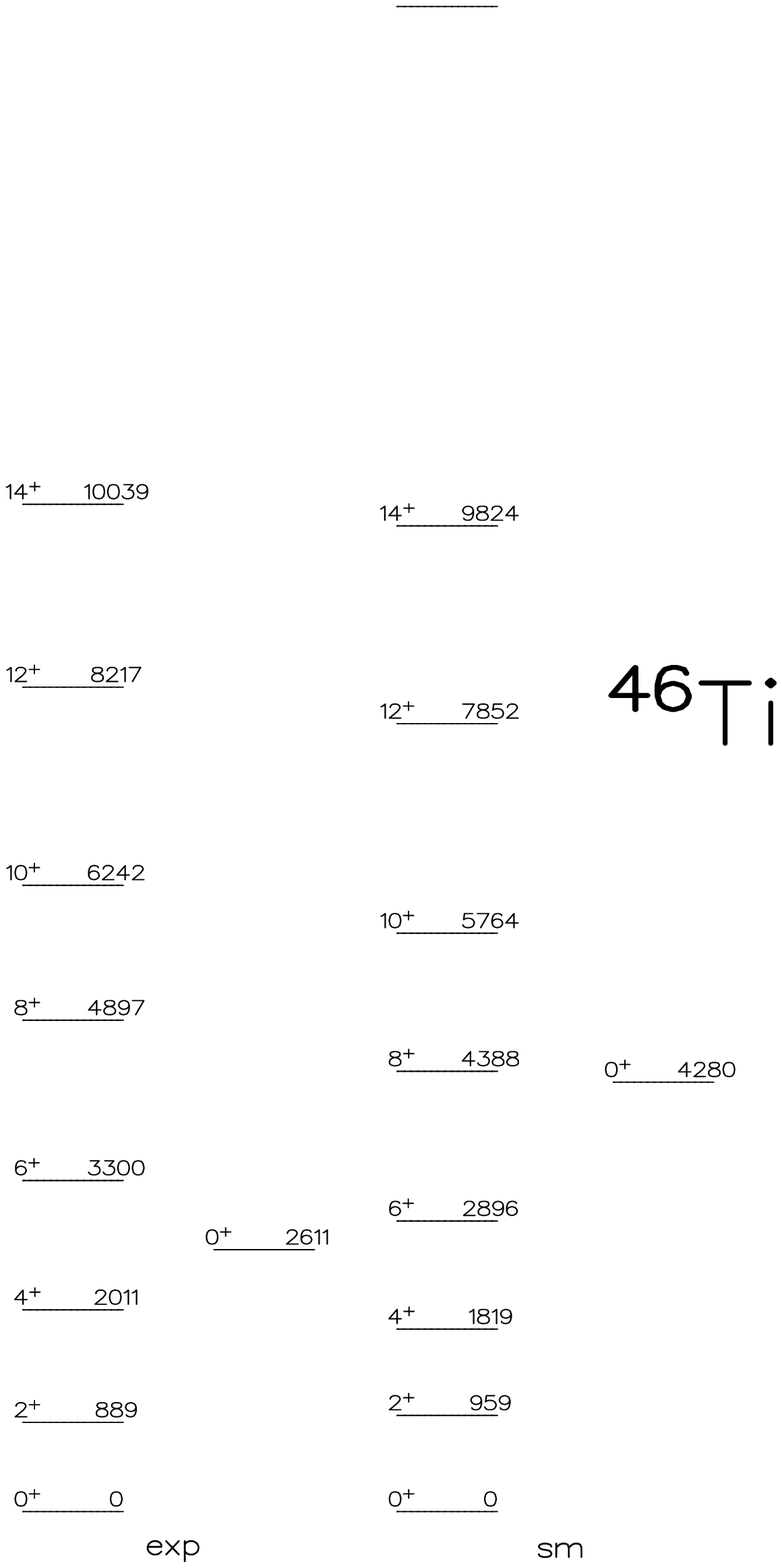,width=6.5cm,angle=0}
 \protect\caption{  Comparison of experimental gs band levels in $^{46}$Ti with SM predictions.}
 \label{fig23}
 \epsfig{file=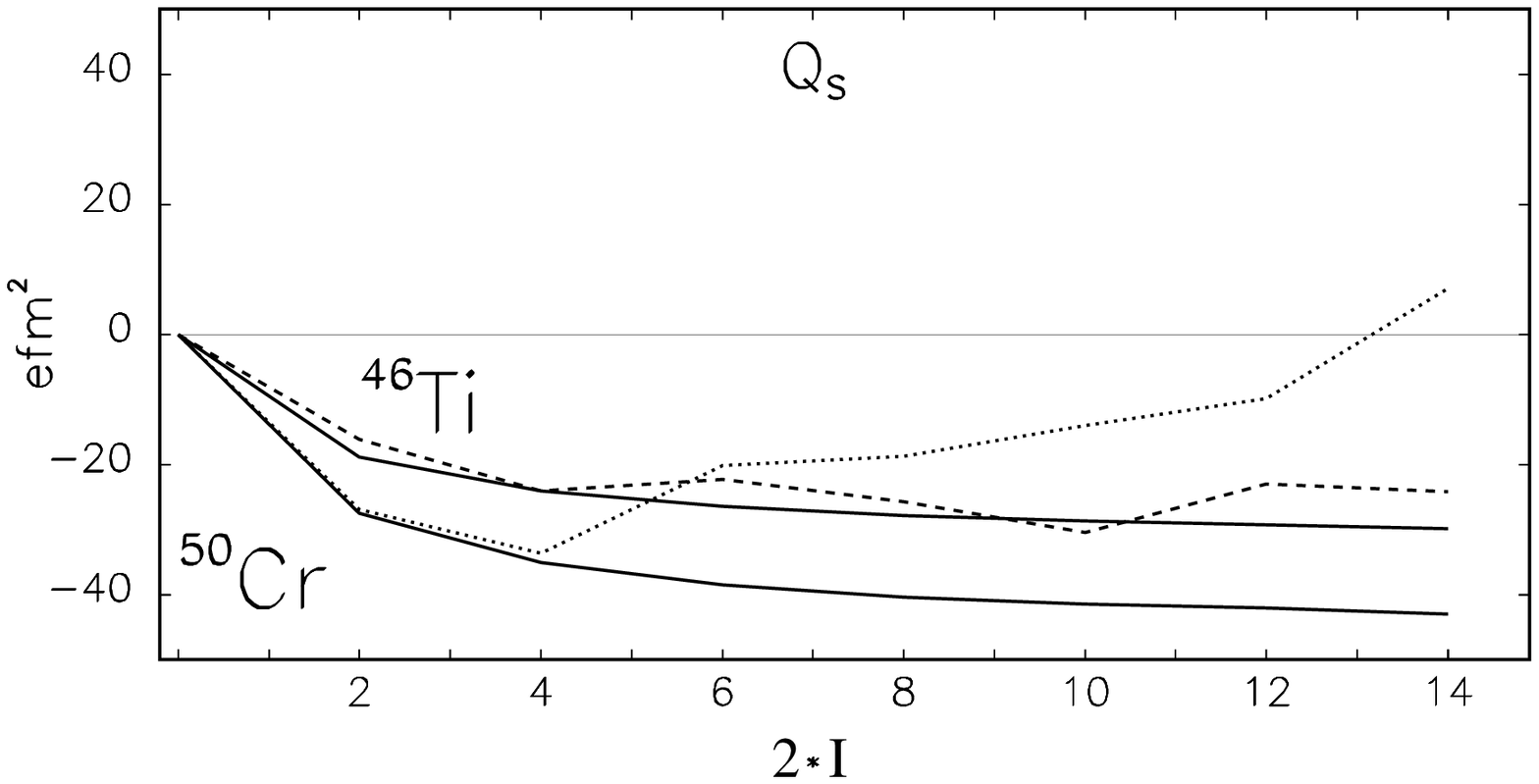,width=8.cm}
 \protect\caption{Comparison between experimental and
                  theoretical  $Q_s$ of $^{50}$Cr ($\beta^*$=0.28) and $^{46}$Ti ($\beta^*$=0.23).}
 \label{fig24}
\end{figure}

Levels not belonging to  previous bands are reported on the rightmost side of Fig.~\ref{fig21}. 
The level density is high both experimentally and theoretically, but the correspondence is good in particular for the two lower 2$^+$ levels.

In Ref.~\cite{Bra50CR} the yrare 2$^+$ level observed at 2924 keV was suggested to be a member of K=1 band, following the antiparallel coupling of two unpaired nucleons. This is qualitatively confirmed by the present SM calculations since the level (calculated at 2811 KeV)  has $Q_s$=10.1 efm$^2$ and it is connected to the yrast 1$^+$ at 3629 keV (calculated at  3539 keV) and to the third 3$^+$ calculated at 3767 keV with $B(E2)$ rates of 241 and 157 efm$^2$, respectively. The yrast 1$^+$ level is estimated to have a remarkable $B(M1)$ value of 0.35 $\mu^2$ towards the gs so that it was observed in inelastic electron  excitation ($e$,$e^{'}$) \cite{Richt}. 

 The positive  $Q_s$ of  29.5 efm$^2$ predicted for the third 2$^+$ level at 3161 keV (calculated at 3000 keV)  makes it a candidate for  the $\gamma$-bandhead. This is confirmed by the very fragmented wave function,  but there should be, two 3$^+$ levels, at 3368 and 3437 keV, which  strongly mix. It looks like as the $K$=2$^+$ mixes strongly with a $K$=3$^+$ band, having  vibrational character too. This complex situation is not represented in Fig.~\ref{fig21} and it is not further discussed, owing to the lack of experimental information.

 In Fig.~\ref{fig23} the experimental gs band of the cross-conjugate nucleus $^{46}$Ti is compared with SM predictions.
 Similarly to $^{44}$Ti, the low spin states of the gs band seem to mix with 2- and 4-hole core configurations, even if at a reduced scale. In fact, if we lift up the SM values by about 400 keV to get  the  8$^+$ levels close, one deduces that the 0$^+$, 2$^+$ and 4$^+$ levels are more bounded by about such value. Moreover the yrare 0$^+$ level is calculated at 4280 keV while it is observed at 2611 keV.
 The experimental $B(E2)$ values for the yrast 2$^+$ and 4$^+$ levels are larger than predicted \cite{Bra46TIPRC}, owing to the relevant contribution of the 2- and 4-hole configurations, while the 6$^+$ level is essentially unaffected as it results from the level scheme.

 In Fig.~\ref{fig24} the  $Q_s$ values in $^{50}$Cr are compared with the ones in its cross--conjugate nucleus $^{46}$Ti. The comparison is made neglecting the  core contribution in $^{44}$Ti. The estimated deformation of $^{46}$Ti is smaller ($\beta^*\simeq$0.23) than in $^{50}$Cr but still relevant  at low spins.
In $^{46}$Ti the change from collective to non--collective regime seems to occur in conjunction with the backbending at $I^\pi=10^+$, similarly to that at $I^\pi=12^+$ in $^{48}$Cr. A major difference  is that there is not a simple relation with a termination in a seniority subspace.
 It seems that the change of regime does not necessarily start at a seniority subspace termination, as previously observed for $^{47}$Ti.

\section{Unnatural parity bands}
  
The shell model description of unnatural parity bands requires 
to consider simultaneously two major shells and, implicitly, severe space 
truncations. Moreover, it is difficult to select a good effective interaction. 
In spite of these severe requirements, reasonable and even satisfactory agreement was achieved in several cases. It is to early for a detailed discussion so that only a brief summary will be presented, limited to odd-$A$ nuclei.
 
Low--lying $I^\pi=3/2^+$ and 1/2$^+$ levels, interpreted as heads of the bands 
having [202]3/2$^+$ and [200]1/2$^+$ Nilsson configurations, were observed in 
$N=Z$+1 nuclei $^{45}$Ti, $^{47}$V, $^{49}$Cr and $^{51}$Mn.
 Such 1-hole configurations are 
selectively populated in pickup reactions: the heads of the $K^\pi$=3/2$^+$ band 
were identified at 293, 260, 1982 and 1817 keV, respectively, while the heads of the 
$K^\pi$=1/2$^+$ band  at 1565, 1660, 2578  and  2276 keV, respectively.
The assignments in $^{51}$Mn are based on the decay scheme and systematics. 

SM calculations were performed mostly for the $K^\pi$=3/2$^+$ band and in this case the $pf$ configuration space was enlarged to include a hole in the 1$d_{3/2}$ orbital. These bands have generally a  deformation different from that of the gs one.
Recently, also the $K^\pi$=1/2$^+$ band in $^{49}$Cr was described, but for this, the extension to the 2$s_{1/2}$ orbital was necessary~\cite{BraUNP}. In fact, the [200]1/2$^+$ Nilsson orbital contains a  large contribution from the spherical 2$s_{1/2}$ orbital. Several effective interactions have been used,  some of them being derived from that used in Ref.~\cite{PovesV} by adjustment of monopole terms.
    
The high--K bands predicted by SM at the lowest energy in such configuration space have $K^\pi$=9/2$^+$, 11/2$^+$, 13/2$^+$ and 15/2$^+$ in $^{45}$Ti, $^{47}$V, $^{49}$Cr and $^{51}$Mn, respectively. 
They are produced by promoting a [202]3/2$^+$ nucleon to the first empty 
orbital and the recoupling of the three unpaired nucleons to the maximum K 
value, obviously, with the necessary mixing of proton and neutron holes in 
order to keep isospin conservation. They can  be also represented as the configurations 
$\pi d_{3/2}^{-1}\otimes^{46}$V(K$=$3,T$=$0), 
$\nu d_{3/2}^{-1}\otimes^{48}$V(K$=$4,T$=$1),
$\pi d_{3/2}^{-1}\otimes^{50}$Mn(K$=$5,T$=$0) and 
$\nu d_{3/2}^{-1}\otimes^{52}$Mn(K$=$6,T$=$1), respectively.

The dipole band $K^\pi$=13/2$^+$ in $^{49}$Cr is yrast and the observed first 
five members  are well reproduced by SM 
calculation~\cite{Bra49CR}. Its band termination is predicted by SM at 33/2$^+$.
 Evidence  was also found of the $K^\pi$=7/2$^+$ partner  band, obtained from an 
antiparallel coupling in the $\nu d_{3/2}^{-1}\otimes^{50}$Mn(I$=$5,T$=$0) 
configuration~\cite{BraUNP}.

Members of the dipole band $K^\pi$=9/2$^+$ in $^{45}$Ti have been also recently 
observed from $I^\pi=17/2^+$ up to the 
termination at $I^\pi=33/2^+$~\cite{Bed}. The reason why members with spin value lower than 17/2$^+$ are  
not observed may be that the 3$^+$ and 5$^+$ terms are rather close to the 
7$^+$ level in $^{46}$V, so that the decay--out is favoured by the transition 
energy.  The  SS reflects that of the $K^\pi$=3$^+$ band in $^{46}$V~\cite{Bra46V}, where it  becomes larger approaching the band termination. For this reason dipole transitions and  the unfavoured signature members were not observed both in the $K$=3 band in $^{46}$V and in the $K^\pi$=9/2$^+$ band in $^{45}$Ti.
  
Unfortunately, non yrast structures are not known in $^{47}$V, so that also 
 the $K^\pi$=11/2$^+$ band was not observed.
 
In $^{51}$Mn the $K^\pi$=15/2$^+$ band was also not observed but the 
experimental situation is intriguing. Recently, a positive parity band   was observed up to 39/2$^+$ \cite{Ek2} and attributed 
to the $\nu g_{9/2}\otimes^{50}$Mn(K$=$5,T$=$0) configuration, which 
is intruder with respect to the so far considered configuration space. 
 The termination  in the configuration space is 
39/2$^+$, in agreement with observation.
Since $^{50}$Mn is prolate, the decoupled [440]1/2$^+$ intruder orbital should be 
considered.
 Assuming the coupling of the band $^{50}$Mn(K$=$5,T$=$0)  with the 9/2$^+$ state of the  decoupled $1g_{9/2}$ band, one would expect a 19/2$^+$ bandhead. Dipole transitions connecting the unfavoured signature levels were not observed above 27/2$^+$. This seems to reflect the fact that unfavoured signature members were not observed in $^{50}$Mn above I=9, because of the large SS \cite{Sven}. The situation is similar to that discussed for the $K$=9/2 band in $^{45}$Ti.
 
 The reason why  the predicted partner quadrupole band 
$\pi g_{9/2}^{-1}\otimes^{50}$Cr(K$=$0,T$=$1), terminating at $I^\pi=37/2^+$, 
is not observed, may be that its bandhead  9/2$^+$ is expected at similar energies as the 19/2$^+$ one  and it is thus  largely non yrast. There is also an explanation of the reason why  the predicted ``extruder'' $K^\pi$=15/2$^+$ $\nu d_{3/2}^{-1}\otimes^{52}$Mn(K$=$6,T$=$1) band is not observed. An 
estimate of its excitation energy is made considering that the IAS level of 
the 6$^+$ gs level of $^{52}$Mn in $^{52}$Fe is at 5655 KeV. Adding this value 
to the excitation energy of the yrast 3/2$^+$ in $^{51}$Mn of 1817 keV one gets 
7472 keV, which is similar to the excitation energy of the lowest terms of the 
observed band. Its termination is 25/2$^+$, well below 39/2$^+$, so that the flux generated at high spin is efficiently captured by the ``intruder'' band, while little goes into the ``extruder'' one.

Large ($\tau,\alpha$) spectroscopic factor with $\ell_n$=0 and $\ell_n$=2 were observed 
in $^{53}$Fe for the levels at 3400 and 2967 keV, respectively, so that they 
are identified as the 3/2$^+$ and 1/2$^+$ states dominated by the 1$d_{3/2}$ 
and 2$s_{1/2}$ orbitals, respectively, which appear to be inverted.

As mentioned, unnatural parity bands were observed also in several
even--even and odd--odd nuclei and were described by lifting a [202]3/2$^+$ 
nucleon to the $pf$ space \cite{PovesV}. Satisfactory agreement was achieved in even-even $^{48}$Cr, $^{50}$Cr and $^{46}$Ti~\cite{Bra48CR50,BraGASP,BraUNP}, as well as in odd-odd nuclei $^{46}$V, $^{48}$V~\cite{Bra46V,Bra48V}.

\section{Conclusions}
 
A critical review is presented of the collective phenomena predicted by SM in several odd--$A$ and even--even nuclei of the 1$f_{7/2}$ shell, for natural parity 
levels up to the band termination in a $f_{7/2}^n$ space.
Deformation alignment (strong coupling) occurs at low excitation energy in odd--$A$ nuclei 
in the second half of the shell, while evidence of rotational alignment is 
limited to  nuclei in the first half of the shell.

Evidence of vibrational $\gamma$--bands is found in both $^{48}$Cr and $^{52}$Fe.
The mixing of the gs band of $^{52}$Fe with sidebands is better outlined.


The Nilsson diagram predicts correctly the lowest sidebands in odd--A nuclei.  
The lowest 3--qp sidebands in $^{49}$Cr and $^{51}$Mn 
have $K^\pi$=13/2$^-$ and 17/2$^-$ respectively, as they are described by 
the excitation of one nucleon from the [321]3/2$^-$ to [312]5/2$^-$ 
orbital or from the [312]5/2$^-$ to the [303]7/2$^-$ orbital, respectively.
Only in the second case a clear crossing with the gs band occurs,
while the backbending of $^{49}$Cr gs band at 19/2$^-$ stems probably from a smooth band 
termination in the $v=$3 subspace. This conclusion shows that in general a 
backbending is not necessarily caused by bandcrossing and that the statement 
based on PSM calculations that backbendings in $^{49}$Cr and $^{51}$Mn are both 
caused by the crossing with a 3-qp $K^\pi$=7/2$^-$ band~\cite{Vel} is untenable. 

The collective rotation is  damaged in N$=$Z$+$1 nuclei $^{47}$V, 
$^{49}$Cr and $^{51}$Mn above $I=$13/2, indicating a change of regime.

 
Similarly to $^{49}$Cr, the understanding of the backbending at 
$I=$12 in $^{48}$Cr, cannot be achieved in the frame of the standard mean field 
approximation. In CHFB, only a treatment sensible to details of the effective 
two body interaction, can predict a backbending   at 
$I^\pi=12^+$ \cite{Tanaka}.
The evolution of $Q_s$ points to a sudden change of regime from collective
to nearly spherical, in absence of any bandcrossing. Experimental data and SM 
calculations exclude, in fact, a crossing of the $^{48}$Cr gs band with a 
 4--qp  K$=$2 band, which served in the PSM to explain the experimental backbending \cite{Ha}. 

What seems principally to occur is the competition between the pairing and the 
quadrupole terms of the nucleon--nucleon interaction. With increasing spin, 
the number of interacting particles in the $2p_{3/2}$ orbital decreases. Consequently the rotational collectivity induced by the quadrupole term also decreases. The pairing interaction 
becomes, therefore, relatively more important, leading in several cases to a rapid  
change from collective to non--collective regime. This change, often 
correlated with a termination in a seniority subspace, occurs in different 
manners, not yet  understood. 
      
The gs bands in nuclei lying in the second half of the 1$f_{7/2}$ shell, as 
$^{51}$Mn, $^{51}$Cr, $^{52}$Fe and $^{53}$Fe, become prolate non--collective 
approaching the termination, but the states are not pure $1f_{7/2}^n$, as commonly believed, as the other orbitals of the $pf$ configuration space contribute to a large extent, increasing the state deformation.
 
A  review of  non natural parity structures is also presented.   

The 1$f_{7/2}$ region is a unique testing bench for the study of the origin 
of nuclear quadrupole deformation. There are more mechanisms of generating deformation but 
what occurs in the 1$f_{7/2}$ shell is a well understood case, where most of 
the game is played by the orbitals 1$f_{7/2}$ and 2$p_{3/2}$.
  SM is a precise microscopic probe of any particle--alignment mechanisms.
It predicts accurate observables,  which react to most structural effects,
where other models may fail. Since SM provides very reliable values for the 
observables in the  1$f_{7/2}$ shell, nuclear models based on the deformed meanfield 
approximation should calibrate their predictions on the SM ones.
In particular, those models which assume a fixed deformation cannot be 
considered reliable in this region. While the 1$f_{7/2}$ region may appear as 
a garden for SM calculations, it is rather a mined field for meanfield models.
   

The general picture presented appears to be well grounded, but new experimental data are required to confirm  the coexistence of rotational and vibrational degrees of freedom predicted by SM.  For this purpose full spectroscopy using light-projectiles induced reactions appears to be the best chance in order to measure accurately the properties of some non-yrast levels of rather low spin.


\acknowledgments
  
This theoretical survey follows a long experimental work performed at LNL.
Several discussions with coworkers are acknowledged.


\begin{thebibliography}{200}

\bibitem{Caur1}
 E.~Caurier, A.P.~Zuker, A.~Poves and G.~Martinez--Pinedo, Phys. Rev. C {\bf 50}, 225 (1994);

\bibitem{Caur2} 
 E.~Caurier {\it et al.}, Phys. Rev. Lett. {\bf 75},  2466 (1995);

\bibitem{Mart1} 
 G.~Martinez--Pinedo, A.~Poves, L.M.~Robledo, E.~Caurier, F.~Novacki and A.P.~Zuker, Phys. Rev.C {\bf 54}, R2150 (1996);

\bibitem{Mart2} 
 G.~Martinez--Pinedo, A.P.~Zuker, A.~Poves and E.~Caurier, Phys. Rev. C {\bf 55},187 (1997);

\bibitem{Lenz1} 
 S.M.~Lenzi {\it et al.}, Z. Phys. {\bf A354}, 117 (1996);

\bibitem{Lenz2}  
 S.M.~Lenzi {\it et al.}, Phys. Rev. C {\bf 56}, 1313 (1997);

\bibitem{Bra48CR50} 
 F.~Brandolini {\it et al.}, Nucl. Phys. {\bf A642}, 387 (1998);

\bibitem{BraSev} 
 F.~Brandolini {\it et al., Experimental Nuclear Physics in Europe,}
Seville, Spain, 1999. Editors B. Rubio, M. Lozano and W. Gelletly,
AIP Conf. Proceedings 495, 1999, pag. 189;

\bibitem{Bra50CR} 
 F.~Brandolini {\it et al.}, Phys. Rev. C {\bf 66},  021302 (2002);

\bibitem{Lenz3} 
 S.M.~Lenzi {\it et al.}, Phys. Rev. Lett. {\bf 87}, 122501 (2001) 

\bibitem{Tanaka} 
 T.~Tanaka, K.~Iwasawa and F.~Sakata, Phys. Rev. C {\bf 58}, 2765 (1998);

\bibitem{Juo2} 
 A.~Juodagalvis, I.~Ragnarsson and S.~Aberg, Phys. Lett. {\bf B477}, 66 (2000);

\bibitem{Ha} 
 K.~Hara, Y.~Sun and T.~Mizusaki, Phys. Rev. Lett. {\bf 83}, 1922 (1999);

\bibitem{Juo1} 
 A.~Juodagalvis and S.~Aberg, Phys. Lett. {\bf B428}, 227 (1998);

\bibitem{Bra49CR} 
 F.~Brandolini {\it et al.}, Phys. Rev. C {\bf 60}, R041305 (1999);

\bibitem{Bra47V49CR} 
 F.~Brandolini {\it et al.}, Nucl. Phys. {\bf A693}, 571 (2001);

\bibitem{BraUNP} 
 F.~Brandolini {\it et al.}, to be published;

\bibitem{Caurlast} 
 E.~Caurier , G.~Martinez--Pinedo, F.~Novacki, A.~Poves and A.P.~Zuker, arXiv:nucl-th/0402046 (2004);
 
 
\bibitem{KB3G} 
 A.~Poves,J.~Sanchez-Solano, E.~Caurier and  F.~Novacki, Nucl. Phys. {\bf A694}, 157 (2001);



\bibitem{Bra46V} 
 F.~Brandolini {\it et al.}, Phys. Rev. C {\bf 64}, 044307 (2001);


\bibitem{Bra48V} F.~Brandolini {\it et al.}, Phys. Rev. C {\bf 66}, 024304 (2002);



\bibitem{FPD6} 
 W.A.~Richter, M.G. van der Merwe, R.E.~Julies and B.A.~Brown, Nucl. Phys. {\bf A523}, 325 (1991);

\bibitem{Ots} 
 T.~Otsuka, M.~Honma and T.~Mitsuzaki, Phys. Rev. Lett. {\bf 81}, 1588 (1998);

\bibitem{Hon} 
 M.~Honma, T.~Otsuka, B.A.~Brown and T.~Mizusaki, Phys. Rev. C {\bf 69}, 034335 (2004);

\bibitem{Hase} 
 M.~Hasegawa and K.~Kaneko, Phys. Rev. C {\bf 59}, 1449 (1999);
 
 
\bibitem{ANT1}
 E.~Caurier, Shell Model code ANTOINE, IRES, Strasbourg 1989-2002;

\bibitem{ANT2}
 E.~Caurier and F.~Novacki, Acta Physica Polonica {\bf 30}, 705 (1999);


\bibitem{QSU3} 
 A.~Zuker, J. Retamosa, A. Poves and E. Caurier, Phys. Rev. C {\bf 52} R1742 (1995);

\bibitem{BM} 
 A.~Bohr and B.R.~Mottelson, Nuclear Structure (Benjamin, New York,1975) 
 Vol. 2, pag.45.

\bibitem{Lo} 
 K.E.G.~L\"obner, M.~Vetter and V.~H\"onig, Nuclear Data Tables {\bf 7}, 
 495 (1970);

\bibitem{Rag} 
 S.E.~Larson, G.~Leander and I.~Ragnarsson, Nucl. Phys. {\bf A307},189 (1978);

\bibitem{Will} 
 S.J.~Williams {\it et al.}, Phys. Rev. C {\bf 68}, 011301 (2003);

\bibitem{Ger}  
 W.J.~Gerace and A.M.~Green, Nucl. Phys. {\bf 113}, 641 (1968);
 
 
\bibitem{Esp} 
 J.M.~Espino {\it et al.}, LNL Annual Report 2001, pag 10;


\bibitem{RagNS} 
 I.~Ragnarsson, V.P. Jansen, D.B. Fossan, N.C.~Schmeing and R.~Wadsworth, Phys. Rev. Lett. {\bf 74}, 3935 (1995);
 
 
\bibitem{Guo} 
 B.G.~Dong and H.C.~Guo, Eur. Phys. J. {\bf A17}, 25 (2003).
 

\bibitem{Aberg} 
 S.~Aberg, Nucl. Phys. {\bf A306}, 89 (1978);

\bibitem{Sh} 
 J.A.~Sheik, D.D.~Warner and P. van Isacker, Phys. Lett. {\bf B443}, 1 (1998);
 
 \bibitem{Ur} 
 C.A.~Ur {\it et al.}, Phys. Rev. C {\bf 58}, 3163 (1998);
 
  
\bibitem{Beng} 
 B.~Bengtsson  and S.~Frauendorf, Nucl. Phys. {\bf 327}, 139 (1979);

 
\bibitem{Rusu} 
 C.~Rusu {\it et al.}, Phys. Rev. C {\bf 69}, 024307 (2004);


\bibitem{Cam} 
 J.A.~Cameron {\it et al.}, Pys.Rev. C {\bf 58}, 808 (1998);
 


\bibitem{Ek2} 
 J.~Ekman {\it et al.}, Phys. Rev. C {\bf 70}, 014306 (2004);

 
\bibitem{Ek1} 
 J.~Ekman {\it et al.}, Eur. Phys. J. {\bf A9}, 13 (2000);
 

\bibitem{Bent} 
 M.A.~Bentley {\it et al.}, Pys.Rev.C {\bf 62}, 051303R (2000);

 
\bibitem{BraGASP} 
 F.~Brandolini, Eur. Phys. J. {\bf A20}, 139 (2004);
 

\bibitem{Jess} 
 K.~Jessen {\it et al.}, Phys. Rev. C {\bf 68}, 047302 (2003);


\bibitem{Ur2}  C.A.~Ur, Eur. Phys. J. {\bf A20}, 113 (2004);
 
 
\bibitem{OL} C.D. O'Leary et al., Phys. Rev. C {\bf 61}, 064314 (2000);

\bibitem{Richt} A. Willis et al., Nucl. Phys. {\bf A499}, 367 (1989).

 \bibitem{Bra46TIPRC} F. Brandolini et al.,  submitted to Phys. Rev. C;

 
\bibitem{PovesV} A. Poves and J. Sanchez-Solano, Phys. Rev. C 58, 179 (1998);

\bibitem{Bed} 
 P.~Bednarczyk {\it et al.},  Eur. Phys. J. {\bf A20},45 (2004);

\bibitem{Sven} C.E. Svenson et al., Phys. Rev C {\bf 58} R2621 (1998):

\bibitem{Vel} 
 V.~Velasquez, J.G.~Hirsch, Y.~Sun, Nucl. Phys. {\bf A686}, 129 (2001);
 



\end{thebibliography}
\end{document}